\RequirePackage{fixltx2e}
\documentclass[aip,jcp,reprint,floatfix,twocolumn]{revtex4-1}
\usepackage{graphicx} 
\usepackage{color}
\usepackage{pagecolor,lipsum}
\usepackage{dcolumn}
\usepackage{amsmath,amsfonts} %
\usepackage{braket} %
\usepackage[acronym]{glossaries} %
\usepackage[breaklinks, %
colorlinks, %
linkcolor=blue, %
citecolor=blue, %
urlcolor=blue]{hyperref}

\newcommand{\eq}[1]{Eq.~(\ref{#1})} %
\newcommand{\bea}{\begin{eqnarray}}
\newcommand{\eea}{\end{eqnarray}}
\newcommand{\mat}[1]{\ensuremath{\boldsymbol{#1}}}

\newcommand{\op}[1]{\ensuremath{\hat{#1}}}

\renewcommand{\braket}[2]{\ensuremath{\langle #1|#2\rangle}}

\newcommand{\braOket}[3]{\ensuremath{\langle #1|#2|#3\rangle}}

\renewcommand{\Re}{\operatorname{Re}}

\renewcommand{\L}{\mathcal{L}}

\newcommand{\e}[1]{\ensuremath{\, \mathrm{e}^{#1}}}

\newcommand\norm[1]{\lVert#1\rVert}

\newcommand{\iu}{{i\mkern1mu}}
\newcommand{\tr}{\ensuremath{\, \mathrm{Tr}}}
\newcommand{\numeq}{\simeq} 
\newcommand{\aU}{\breve{\boldsymbol U}}
\newcommand{\aUs}{\breve{\boldsymbol u}}
\newcommand{\Regone}{\emph{Reg1}}
\newcommand{\Regtwo}{\emph{Reg2}}
\newcommand{\Var}{\emph{Var}}

\newacronym{SE}{SE}{Schr\"odinger equation}
\newacronym[\glslongpluralkey=degrees of freedom]{DOF}{DOF}{degree of freedom}
\newacronym{TDVP}{TDVP}{time-dependent variational principle}
\newacronym{MCTDH}{MCTDH}{multi configuration time-dependent Hartree}
\newacronym{GMCTDH}{G-MCTDH}{Gaussian multi configuration time-dependent Hartree}
\newacronym{vMCG}{vMCG}{variational multi configuration Gaussian}
\newacronym{AIMS}{AIMS}{ab initio multiple spawning}
\newacronym{MCE}{MCE}{multi configuration Ehrenfest}
\newacronym[\glslongpluralkey=equations of motion]{EOM}{EOM}{equation of motion}
\newacronym{TDBF}{TDBF}{time-dependent basis function}

\begin{document}

\author{Lo{\"i}c Joubert-Doriol}
\affiliation{Univ Gustave Eiffel, Univ Paris Est Creteil, CNRS, UMR 8208, MSME, F-77454 Marne-la-Vall\'ee, France} %
\email{loic.joubert-doriol@univ-eiffel.fr}

\title{A variational approach for linearly dependent moving bases in quantum dynamics: application to Gaussian functions}

\begin{abstract}
In this paper, we present a variational treatment of the linear dependence for a non-orthogonal time-dependent basis set in solving the Schr\"odinger equation.
The method is based on: i) the definition of a linearly independent working space, and ii) a variational construction of the propagator over finite time-steps.
The second point allows the method to properly account for changes in the dimensionality of the working space along the time evolution.
In particular, the time evolution is represented by a semi-unitary transformation.
Tests are done on a quartic double-well potential with Gaussian basis function whose centers evolve according to classical equations of motion.
We show that the resulting dynamics converges to the exact one and is unitary by construction.
\end{abstract}

\date{\today}

\maketitle

\section{Introduction}
\label{sec:introduction}

Describing nature at the microscopic scale often requires solving the \gls{SE}.
However, this equation cannot generally be solved exactly, and perturbative or variational approximations are necessary.
Variational approaches are particularly appealing thanks to their flexibility in describing the systems of interest and because they do not rely on the definition of an appropriate zeroth order Hamiltonian.
A particularly powerful approach consists in expressing the approximate \gls{SE} solution in a set of \glspl{TDBF} that represent best the wave packet at any time along the dynamics.
In the context of molecular systems, application of the \gls{TDVP}~\cite{Frenkel:1934/Book,Dirac:1958/Book,Mclachlan:1964/mp/39,Kramer:1981/Book,Hackl:2020/sp/048} resulted in some of the most used quantum variational methods for electronic dynamics~\cite{Caillat:2005/pra/012712,Sasmal:2020/jcp/154110} or nuclear dynamics, such as the \gls{MCTDH} method.~\cite{Beck:2000/pr/1,Meyer:2009/book} 
Of particular interest are variations that employ Gaussian \glspl{TDBF} evolving according to the \gls{TDVP}: \gls{GMCTDH} and \gls{vMCG},~\cite{Burghardt:1999/jcp/2927,Worth:2004/fd/307,Richings:2015/irpc/269,Joubert:2018/jpca/6031,Worth:2020/cpc/107040} or non-variational: \gls{AIMS} and \gls{MCE}.~\cite{Bennun:2000/jpca/5161,Ben-Nun:2002/Book,Makhov:2017/cp/200,Curchod:2018/cr/3305,Lassmann:2021/jcp/211106}

A variational treatment of the time-dependent basis often involves a non-linear parameterization (e.g. \gls{MCTDH} or \gls{vMCG}), which results in the coupled \glspl{EOM} of the \glspl{TDBF}.
Solving these equations is generally a difficult task since it requires inverting potentially large matrices.
To reduce computational cost, one can decouple the \glspl{TDBF}.
Then, the time evolution of the basis is not variational anymore, and the solution to the \gls{SE} is not optimal but is still variational.
This decoupling is often employed for the time evolution of \glspl{TDBF} associated with nuclear \glspl{DOF}, represented by Gaussians~\cite{Heller:1975/jcp/1544,Heller:1976/jcp/63,Lasser:2020/an/229,Garashchuk:2021/Book} or coherent states~\cite{Bargmann:1971/rmp/221,Shalashilin:2001/jcp/5367,Werther:2021/irpc/81} whose centers follow classical trajectories: for example using Erhenfest~\cite{Shalashilin:2009/jcp/244101,Makhov:2017/cp/200,Makhov:2022/jpc/025001}, or Born-Oppenheimer trajectories.~\cite{Bennun:2000/jpca/5161,Yu:2020/jacs/20680,Ibele:2021/jcp/174119}
The independent \glspl{TDBF} approach projects the \gls{SE} on a time-dependent (working) space spanned by the \glspl{TDBF}.
Then, the wavefunction reads
\bea\label{eq:WFdef}
\ket{\Psi(t)} & = & \sum_{k=1}^{N_g} \ket{g_k(t)} C_k(t),
\eea
where, $\{g_k(t); k=1\dots,N_g\}$ are the $N_g$ \glspl{TDBF} (which are not limited to Gaussian functions), and $C_k(t)$ are complex time-dependent coefficients.
Another advantage of utilizing independent \glspl{TDBF} is the possibility for efficient parallelization.~\cite{Shalashilin:2009/jcp/244101,Curchod:2018/cr/3305}
These independent \glspl{TDBF} are not variationally optimal for the solution of the \gls{SE}, but this drawback can be compensated by propagating a large number of \glspl{TDBF} to reach completeness.
Once the time evolution of the \glspl{TDBF} is known, the \gls{SE} is approximately solved after projection on the time-dependent basis.
This projector reads
\bea\label{eq:Porjdef}
\op P_g (t) & = & \sum_{k=1}^{N_g} \sum_{l=1}^{N_g} \ket{g_k (t)}[\mat S^{-1}]_{kl}\bra{g_l (t)}, \label{eq:fullproj}
\eea
where $\mat S$ is the overlap matrix defined by
\bea
S_{kl} (t) & = & \braket{g_k (t)}{g_l (t)}. \label{eq:S}
\eea

\Acrlongpl{TDBF} commonly overlap largely with each other.
This can create a near-linear dependence of the basis, which is in fact a numerical linear dependence that result in a singular overlap matrix $\mat S (t)$.
Since the inverse of the overlap matrix appears in the definition of the projector \eq{eq:fullproj}, singularities in the overlap indicates an ill-defined working space.
A signature of this problem arises in the \gls{EOM} for the coefficients $\mat C (t)$ where the overlap matrix also appears,~\cite{Richings:2015/irpc/269,Curchod:2018/cr/3305,Makhov:2017/cp/200}
\bea\label{eq:EOMSC}
\mat S (t) \dot{\mat C} (t) & = & - [ \iu\mat H (t) + \mat\tau (t) ] \mat C (t), \label{eq:Cdot}
\eea
where the dot symbolizes the time-derivative and
\bea
\tau_{kl} & = & \braket{g_k (t)}{\dot{g}_l (t)}, \label{eq:tau}\\
H_{kl} & = & \braOket{g_k (t)}{\op H}{g_l (t)}. \label{eq:H}
\eea
We employ atomic units in \eq{eq:Cdot} and in the rest of the paper.
From \eq{eq:Cdot}, we observe that a singular overlap leads to an ill-defined set of equations for the coefficients' evolution.~\cite{Sawada:1985/jcp/3009,Kay:1989/cp/165,Burghardt:1999/jcp/2927,Habershon:2012/jcp/014109,Richings:2015/irpc/269,Hackl:2020/sp/048}

While the linear dependence of the basis is more pronounced in the case of independent \glspl{TDBF}, we need to mention that a similar problem also occurs in methods employing variational \glspl{TDBF} (e.g. \gls{MCTDH} and \gls{vMCG}).
Even, if an equivalent to \eq{eq:Cdot} also exists for these approaches, the problem is less significant since optimal evolution is assumed to diminish \glspl{TDBF} overlaps in order to improve completeness, which also reduces the linear dependence.
However, another \gls{EOM}, used to solve for the time-evolution of the \glspl{TDBF}, shows a similar problem. 
Indeed, an overlap of variations along non-linear parameters of the wave function (not the basis overlap) must be inverted to solve this \gls{EOM}, but can also become singular.~\cite{Hackl:2020/sp/048}
This occurs, for example, in the \gls{MCTDH} method,~\cite{Meyer:2009/book,Manthe:2015/jcp/244109,Lubich:2015/amre/311,Meyer:2018/jcp/124105} or in \gls{GMCTDH} and \gls{vMCG}.~\cite{Polyak:2015/jcp/084121}
Thus, investigating the linear dependence in the case of independent \glspl{TDBF} is a first step to the more general problem depicted here.

To remedy the problem of a singular $\mat S$, one can act on the time-evolution of the \glspl{TDBF}.
A possible approach is then to reintroduce some correlation between the \gls{TDBF} so that they overlap less significantly.
The optimal approach employing variational basis that is mentioned in the previous paragraph does exactly that.
We can also mention methods that re-spawn the basis at each time-step such as the ``matching pursuit'' algorithm~\cite{Wu:2003/jcp/6720} or the ``Basis Expansion Leaping''.~\cite{Koch:2013/prl/263202}
However, in the present work, we assume that the time evolution of the \glspl{TDBF} is given, and we rather want to find an approach to handle the linear dependence and solve for the \gls{SE} in this basis.

Another approach is to regularize $\mat S$ using an approximate but invertible matrix, which is then used to approximate $\mat S^{-1}$.
This approach is employed in \gls{MCTDH},~\cite{Meyer:1990/pcl/73,Manthe:1992/jcp/3199} \gls{vMCG},~\cite{Polyak:2015/jcp/084121,Richings:2015/irpc/269} or \gls{AIMS}.~\cite{Martinez:1996/jcp/2847}
Similarly, one can replace $\mat S^{-1}$ with its Moore-Penrose pseudo-inverse.~\cite{Kay:1989/cp/165}
We can also mention the iterative construction of an approximate $\mat S^{-1}$ using, for example, the Hotelling’s method.~\cite{MauritzAndersson:2001/jcp/1158}
Nevertheless, one effect of approximating the inverse of $\mat S$ is that it modifies \eq{eq:Cdot}.
Since \eq{eq:Cdot} is variational, departing from it can result in a non-variational solution,~\cite{Conte:2010/m2an/759} which can prevent convergence.
Hence, we would like to avoid a procedure that modifies \eq{eq:Cdot}.

To eliminate linear dependence, we can reduce the number of \glspl{TDBF} to a subset of all the functions at our disposal.
This is the motivation behind the ideas presented in 
references~\citenum{Habershon:2012/jcp/014109,Richings:2015/irpc/269,Werther:2020/prb/174315} for Gaussian \glspl{TDBF}.
We note that such an approach can efficiently be combined with creation of new basis functions when needed to ensure low error in the \gls{SE} solution.
The creation of new basis functions can either be based on some local data along a single \gls{TDBF} time evolution,~\cite{Bennun:2000/jpca/5161,Makhov:2014/jcp/054110} or based on the full information of the total wavefunction, either perturbatively~\cite{Izmaylov:2013/jcp/104115} or by minimization of the error.~\cite{Izmaylov:2017/jpcl/1793,Martinazzo:2020/prl/150601,Mendive-Tapia:2020/jcp/234114}
The main difficulty comes from discontinuities introduced with the reduction of the working space.
Furthermore, it seems that removing \emph{entire \glspl{TDBF} at a time} to build a linearly independent space is not optimal in the sense that there other larger dimensional linearly independent spaces can be obtained using \emph{linear combinations of the \glspl{TDBF}}, which could improve convergence.
This is the route we follow.

In this paper, we elaborate on these ideas by developing a new approach where we define, based on the given overlapping \glspl{TDBF}, a smaller dimensional linearly independent basis that we can use as a working space to solve the \gls{SE}.
Since the overlap matrix evolves in time, the size of this working subspace varies in time.
To account for this change in dimensionality, we apply a variational approach that is different from the usual \gls{TDVP} in the sense that it is applied for finite time-steps.
We then test and compare this approach against two approaches that regularize \eq{eq:Cdot} on a simple model designed to intensify linear dependence of the \glspl{TDBF}: a one-dimensional double-well, and utilizing a basis of moving Gaussians with frozen width.

The rest of the paper is organized as follows.
Section~\ref{sec:theory} describes the new method.
Computational details, including the model description, are given Sec.~\ref{sec:details}.
Section~\ref{sec:results} presents the results and the related discussion.
We finally conclude with Sec.~\ref{sec:conclusion}.

\section{Description of the method}
\label{sec:theory}

Our approach is based on three ideas developed in the three following subsections.
In subsection~\ref{sub:Wspace}, we construct a well-defined linearly independent subspace to be used as a proper working space to apply the variational principle.
Due to possible changes in the dimensionality of the working space along the time evolution, we define a variational approach over a finite time-step to avoid discontinuities in subsection~\ref{sub:newTDVP}.
A second variational treatment allows us to define unitary time-reversible propagators in subsection~\ref{sub:opt}.

For the sake of compactness we will introduce two specific notations in the rest of the document:
i) we replace the explicit time-dependence by subscripts to the corresponding matrix and vector quantities, e.g. $\mat C_1\equiv\mat C(t_1)$ or $\mat H_2\equiv\mat H(t_2)$ (the time-dependence is simply dropped when the position in time is not essential and no subscripts are given), and ii) we define vectors of basis functions as $\mat g^t\equiv(\ket{g_1}\,\ket{g_2}\,\dots\ket{g_{N_g}})$.
With the introduced notations we have that $\ket{\Psi(t_1)}\equiv\ket{\Psi_1}=\mat g_1^t\mat C_1$ or $\mat\tau(t_1)\equiv\mat\tau_1=\mat g_1^*\dot{\mat g}_1^t$.

\subsection{Definition of the working space}
\label{sub:Wspace}

Our starting point is the working space build from the set of \glspl{TDBF} $\{g_k\}$ at a given time.
We need to isolate $M$ linearly independent states from this basis.
While the working space is unique, its construction is not.
We utilize the eigendecomposition of $\mat S$ (singular value decomposition or Gram–Schmidt process could also be used).
The resulting eigenvalues can be grouped into $M$ non-zero positive eigenvalues $\{s_k;k=1,\dots,M\}$ and $N_g-M$ eigenvalues $\{\zeta_k;k=1,\dots,N_g-M\}$ that are numerically zero in the sense that $\zeta_k<\varepsilon_S$ where $\varepsilon_S$ is a threshold to be defined later.
Eigenvectors with non-zero eigenvalues are represented as a rectangular matrix $\mat V$, while the matrix $\mat W$ contains the remaining ones:
\bea\label{eq:Seigdecomp}
\mat S
& = & (\begin{matrix} \mat V \mat W \end{matrix}) \begin{pmatrix} \mat s & \mat 0 \\ \mat 0 & \mat \zeta \end{pmatrix} (\begin{matrix} \mat V \mat W \end{matrix})^\dagger
\numeq \mat V\mat s\mat V^\dagger.
\eea
The symbol ``$\numeq$'' indicates numerical equality in \eq{eq:Seigdecomp}.
After discarding the eigenvectors associated with numerically zero eigenvalues ($\mat W$), we obtain a set of linearly independent states $\{\ket{\varphi_k};k=1,\dots,M\}$ defined by
\bea
\mat\varphi^t & = & \mat g^t\mat \Phi, \label{eq:phi}\\
\mat\Phi & = & \mat V\mat s^{-1/2}, \label{eq:Phi}
\eea
and the corresponding projector reads
\bea\label{eq:Pphi}
\op P & = & \mat\varphi^t\mat\varphi^* = \mat g^t\mat V \mat s^{-1}\mat V^\dagger\mat g^*.
\eea
We note that this projector is similar to the one defined in \eq{eq:fullproj} but utilizing the Moore-Penrose pseudo-inversion on the singular overlap matrix.
Using this projector, it is easy to check that the remaining basis function combinations form a numerically null space: $\op P \mat g^t\mat W = \mat g^t\mat V \mat s^{-1}\mat V^\dagger\mat S\mat W \numeq 0$.
Therefore, the following relation holds
\bea
\mat g^t & = & \mat g^t\mat V\mat V^\dagger,
\eea
and, in particular, the wavefunction can be rewritten
\bea\label{eq:Psinewform}
\ket{\Psi} & = & \mat g^t\mat V\mat V^\dagger\mat C = \mat\varphi^t\bar{\mat C},
\eea
where we define
\bea
\bar{\mat C} & = & \mat \Phi^-\mat C, \label{eq:Cb}\\
\mat \Phi^- & = & \mat s^{1/2}\mat V^\dagger. \label{eq:Phim}
\eea
From \eq{eq:Psinewform}, we observe that $\mat W^\dagger\mat C$ does not have an impact on the local (in time) dynamics, and we will impose that $\mat W^\dagger\mat C=\mat 0$, and $\mat V\mat V^\dagger\mat C=\mat C$.

Now that we have a well-defined set of linearly independent states that composes the working space, starting from \eq{eq:Psinewform}, we obtain an equation equivalent to \eq{eq:Cdot} but in this new basis (see App.~\ref{app:proofCbdot})
\bea\label{eq:Cbdot}
\dot{\bar{\mat C}}
& = & -[\bar{\mat\tau}+\iu\bar{\mat H}] \bar{\mat C}
\eea
where we used the following definitions
\bea
\bar{\mat\tau} & = & \mat\varphi^*\dot{\mat\varphi}^t = \mat\Phi^\dagger\mat\tau{\mat\Phi}+\mat\Phi^\dagger\mat S\dot{\mat\Phi}, \label{eq:taub}\\
\bar{\mat H} & = & \mat\varphi^*\op H\mat\varphi^t = \mat\Phi^\dagger\mat H\mat\Phi. \label{eq:Hb}
\eea
An important advantage of \eq{eq:Cbdot} comes from the fact that it uses $\mat s^{-1/2}$ rather than $\mat s^{-1}$.
Since $\mat s^{-1/2}$ is better conditioned, we can employ a significantly lower threshold $\varepsilon_S$.

\subsection{A ``finite-step'' variational approach}
\label{sub:newTDVP}

Due to the time-dependence of the overlap matrix, we have that $\mat S_1\neq\mat S_2$ at different times $t_1$ and $t_2$ ($\op P_1\neq\op P_2$).
Hence, it is possible that the dimensionality of the linearly independent working spaces at $t_1$ is different from the one at $t_2$: $M_1\neq M_2$.
This introduces an additional complexity in the integration of \eq{eq:Cbdot}, which becomes discontinuous.
To avoid this problem, we do not combine the \gls{TDVP} with the \gls{SE}, but we rather directly project the solution of the \gls{SE} for a finite step on the working space.
In this aim, we first define the exact wavefunction at time $t_2$
\bea\label{eq:exactPsi2}
\ket{\Psi^e_2} & = & \op U(t_2,t_1) \ket{\Psi_1},
\eea
where the state at $t_1$, $\ket{\Psi_1}$, is imposed.
We only assume that the operator $\op U(t_2,t_1)$ is unitary, but it can be identified, for example, as the exact propagator of the \gls{SE}: $\op U(t_2,t_1)=\e{-\iu\op H (t_2-t_1)}$.
We then project this equation on the subspace at time $t_2$
\bea
\mat\varphi_2^*\ket{\Psi^e_2} & = & \mat\varphi_2^*\op U(t_2,t_1)\ket{\Psi_1} = \aU_{21}\bar{\mat C_1}, \\
\aU_{21} & = & \mat\varphi_2^*\op U(t_2,t_1)\mat\varphi_1^t. \label{eq:exactpropag}
\eea
The resulting vector $\mat\varphi_2^*\ket{\Psi^e_2}$ is a variational approximation to the solution of \gls{SE} since it converges to the exact solution as the basis grows to completeness.

\subsection{Optimization of the transformation}
\label{sub:opt}

It is clear from \eq{eq:exactpropag} that $\aU_{21}$ is time-reversible but not unitary for a finite basis $\{\varphi_k\}$.
The matrix $\aU_{21}$ can even be rectangular if the dimensions of the working spaces at $t_1$ and $t_2$ are different $M_2\neq M_1$.
We would like to generate the best approximation $\bar{\mat U}_{21}$ from $\aU_{21}$ that is unitary and time-reversible.
As we will see in the next subsection, this is only possible when $M_2=M_1$.
When $M_2\ne M_1$, $\aU_{21}$ is rectangular.
Hence, the expected transformation, $\bar{\mat U}_{21}$, is also a rectangular transformation, and we cannot obtain a unitary approximation $\bar{\mat U}_{21}$ to $\aU_{21}$.
Instead, we generate the best semi-unitary approximation, which satisfies $\bar{\mat U}_{21}\bar{\mat U}_{21}^\dagger=\mat 1_{M_2}$ if $M_1>M_2$ or $\bar{\mat U}_{21}^\dagger\bar{\mat U}_{21}=\mat 1_{M_1}$ if $M_2>M_1$, where $\mat 1_M$ is the $M$-dimensional identity matrix.
Such a semi-unitary transformation belongs to a complex Stiefel manifold,~\footnote{The complex Stiefel manifold is the space of all $k$-dimensional semi-unitary matrices in a $D$-dimensional space with $k<D$.} and we need to apply the variational principle on this manifold.~\cite{Hairer:2006/Book} 
To simplify the procedure, we truncate the larger of the two spaces so that unitarity is recovered in this smaller space.
We should also emphasize that once this truncation is operated, one could obtain the time-evolution by integrating \eq{eq:Cbdot} in this truncated subspace.
The resulting transformation in the ``untruncated'' working space is indeed a semi-unitary matrix, but generally not time-reversible.
The overall procedure minimizes the distance between $\aU_{21}$ and $\bar{\mat U}_{21}$ using the Frobenius norm
\bea
\norm{\bar{\mat U}_{21}-{\aU_{21}}} &=& 
\sqrt{\tr\{(\bar{\mat U}_{21}-{\aU_{21}})^\dagger(\bar{\mat U}_{21}-{\aU_{21}})\}}, \label{eq:Lag1eq2}
\eea
under the constraint that $\bar{\mat U}_{21}$ is unitary if $M_2=M_1$ or semi-unitary otherwise.
This approach is variational and converges to the exact dynamics in the infinite basis limit (see App.~\ref{app:variational}.

\subsubsection{The case $M_2=M_1$}

In this case, the matrix $\aU_{21}$ is square.
The construction of the unitary transformation $\bar{\mat U}_{21}$ is achieved by minimizing $\norm{\bar{\mat U}_{21}-{\aU_{21}}}^2$, given by \eq{eq:Lag1eq2}, under the constraint that $\bar{\mat U}_{21}$ is unitary ($\bar{\mat U}_{21}^\dagger=\bar{\mat U}_{21}^{-1}$).
This minimization is equivalent to maximizing the quantity $\Re[\tr\{\bar{\mat U}_{21}^\dagger{\aU_{21}}\}]$.
This is a Procrustes problem and the known solution is found by using singular value decomposition of $\aU_{21}$~\cite{Gower:2004/Book}
\bea
\aU_{21} & = & \mat X\aUs_{21}\mat Y^\dagger, \label{eq:svd1}\\
\bar{\mat U}_{21} & = & \mat X\mat Y^\dagger, \label{eq:svd2}
\eea
where $\mat X$ and $\mat Y$ are unitary, and $\aUs_{21}$ is a real diagonal semi-positive matrix.
Then, we can generate the coefficients of interest in the basis $\{g_k\}$ as follows
\bea\label{eq:properC2}
\mat C_2 & = & \mat\Phi_2\bar{\mat U}_{21}\mat\Phi_1^- \mat C_1.
\eea
When no linear dependencies are present in the basis (\emph{e.g.} $M_2=M_1=N_g$) the solution matches the usual \gls{TDVP} solution given in \eq{eq:Cdot} as shown in App.~\ref{app:equiv}.

\subsubsection{The cases $M_2>M_1$}

To recover unitarity, we proceed in two steps: i) we first truncate the space at time $t_2$ so that its new dimension after truncation is $M_2'=M_1$, then ii) we apply the same procedure as in the previous subsection.
We operate the truncation variationally by defining the basis which maximizes the overlap between the truncated space with the original one.
This basis is represented by the semi-unitary matrix $\mat R$, which satisfies $\mat R^\dagger\mat R=\mat 1_{M_1}$.
In this case, we want to maximize $\mat\aU_{2'1}=\mat R^\dagger\mat\aU_{21}$.
For this purpose, we define the Lagrangian
\bea\label{eq:LagM2ltM1}
\L & = & \norm{\mat R^\dagger\aU_{21}}^2 + \tr\{\mat\lambda(\mat R^\dagger\mat R-\mat 1_{M_1})\},
\eea
where $\mat\lambda$ is a matrix of Lagrange multipliers to enforce the constraint $\mat R^\dagger\mat R=\mat 1_{M_1}$.
The stationary condition leads to an eigenequation
\bea\label{eq:eigM2ltM1}
(-\aU_{21} \aU_{21}^\dagger) \mat R & = & \mat R \mat\lambda,
\eea
from which the maximum of \eq{eq:LagM2ltM1} is obtained by selecting the $M_1$ eigenvalues of \eq{eq:eigM2ltM1} with the largest magnitude.
The square matrix $\aU_{2'1}=\mat R^\dagger\aU_{21}$ (note the prime in the index indicating that the space at $t_2$ is reduced) is then employed in place of the rectangular $\aU_{21}$ to construct the unitary transformation $\bar{\mat U}_{2'1}$.
We apply the singular value decomposition onto $\aU_{2'1}$ and build $\bar{\mat U}_{2'1}$ as in the previous subsection
\bea
\aU_{2'1} & = & \mat X'\aUs_{2'1}(\mat Y')^\dagger, \\
\bar{\mat U}_{2'1} & = & \mat X'(\mat Y')^\dagger.
\eea
Going back from $M_2'$- to the $M_2$-dimensional space at time $t_2$, we can build $\bar{\mat U}_{21}$ as
\bea\label{eq:U12bar}
\bar{\mat U}_{21} & = & \mat R\mat X'(\mat Y')^\dagger.
\eea
It is now easy to check that $\bar{\mat U}_{21}$ is semi-unitary ($\bar{\mat U}_{21}^\dagger\bar{\mat U}_{21}=\mat 1_{M_1}$).

%
\subsubsection{The cases $M_2<M_1$}

As in the previous subsection, we proceed in two steps: i) we reduce the size of the subspace at $t_1$ from $M_1$ to $M'_1=M_2$, and ii) minimization of the error as in the case $M_2=M_1$.
The dimensional reduction is done by maximizing $\mat\aU_{21'}=\mat\aU_{21}\mat R$, with the constraint that $\mat R^\dagger\mat R=\mat 1_{M_2}$.
We also need the subspace at time $t_1$ to contain the vector $\mat C_1$ and possibly $N_c-1$ other vectors of importance (with $N_c\le M_1$).
We recast this latter requirement as a new constraint $\mat R\mat R^\dagger\mat D=\mat D$, where $\mat D$ contains the $N_c$ important vectors (including $\mat C_1$) already orthonormalized $\mat D^\dagger\mat D=\mat 1_{N_c}$.
To impose this last constraint, we choose the parameterization $\mat R = (\mat D \mat E)$.
Hence, we have to maximize $\aU_{21}\mat E$ under the constraints $\mat E^\dagger\mat E=\mat 1_{(M_2-N_c)}$ and $\mat E^\dagger\mat D=\mat 0$.
For this purpose, we define the Lagrangian
\bea\label{eq:LagM1ltM2}
\L & = & \norm{\aU_{21}\mat E}^2 + \tr\{\mat\lambda(\mat E^\dagger\mat E-\mat 1_{(M_2-N_c)})\} \nonumber\\
&&+ \Re[\tr\{\mat\lambda'(\mat E^\dagger\mat D)\}], 
\eea
where $\mat\lambda$ and $\mat\lambda'$ are matrices of Lagrange multipliers.
Applying the stationary condition to \eq{eq:LagM1ltM2} gives a set of three equations
\bea
\aU_{21}^\dagger\aU_{21}\mat E + \mat E\mat\lambda + \mat D\mat\lambda' & = & \mat 0, \label{eq:statM1ltM2-1}\\
\mat E^\dagger\mat E-\mat 1_{(M_2-N_c)} & = & \mat 0, \label{eq:statM1ltM2-2}\\
\mat E^\dagger\mat D & = & \mat 0. \label{eq:statM1ltM2-3}
\eea
Multipliying \eq{eq:statM1ltM2-1} by $\mat D^\dagger$ on the left, we obtain 
\bea
\mat\lambda' & = & -\mat D^\dagger\aU_{21}^\dagger\aU_{21}\mat E.
\eea
We can then substitute for $\mat \lambda'$ in \eq{eq:statM1ltM2-1} and use \eq{eq:statM1ltM2-3} to obtain the eigenequation
\bea
-(\mat 1_{N_c} - \mat D\mat D^\dagger)\aU_{21}^\dagger\aU_{21}(\mat 1_{M_2} - \mat D\mat D^\dagger)\mat E & = & \mat E\mat\lambda, \label{eq:eigM1ltM2}
\eea
from which the maximum of \eq{eq:LagM1ltM2} is obtained by selecting the $M_2-N_c$ eigenvalues of \eq{eq:eigM1ltM2} with the largest magnitude.
The singular value decomposition of the matrix $\mat\aU_{21'}=\mat\aU_{21}\mat R=\mat X'\mat\aUs_{21'}(\mat Y')^\dagger$ is then employed to construct the closest unitary transformation $\bar{\mat U}_{21'}=\mat X'(\mat Y')^\dagger$.
The time-evolution is then given by the propagator
\bea
\bar{\mat U}_{21} & = & \mat X'(\mat Y')^\dagger\mat R^\dagger.
\eea
Again, $\bar{\mat U}_{21}$ is semi-unitary by construction ($\bar{\mat U}_{21}\bar{\mat U}_{21}^\dagger=\mat 1_{M_2}$), and it satisfies $\bar{\mat U}_{21}^\dagger\bar{\mat U}_{21}\mat D=\mat D$.

\section{Computational details}
\label{sec:details}

\subsection{Regularized methods}

We compare variational the approach presented in Sec.~\ref{sec:theory}, named the ``{\Var}'' approach hereafter, to two approaches where we handle the linear dependence through regularization in order to solve \eq{eq:Cdot}.
In these two other approaches, the singular overlap matrix $\mat S$ is inverted approximately into $\mat S^{-}$.

The first approach employs a Moore-Penrose pseudo-inversion where the inverse of the numerical zeros in the eigenvalues of $\mat S$ are replaced by $0$ similarly to the pseudo-inverse done in \eq{eq:Seigdecomp} so that $\mat S^{-}=\mat V\mat s^{-1}\mat V^\dagger$.
Numerical zeros are defined as being smaller than a threshold $\varepsilon_S$.
We shall name this method  ``{\Regone}''.
While {\Regone} seems equivalent to the {\Var} method, the difference lies in the fact that for \Var, the pseudo-inversion is used to build a well defined working space as a starting point to apply the variational treatment, while in \Regone, the pseudo-inversion is employed \emph{a posteriori}, which can break the variational character of \eq{eq:Cdot}.

The second approach we consider here is often employed in combination with the \gls{MCTDH} method to regularize the reduced density matrix in single particle functions propagation.~\cite{Meyer:1990/pcl/73}
In this approach, numerical zeros in the eigenvalues of $\mat S$, again defined as being smaller than $\varepsilon_S$, are replaced by a small number such that the approximate inverse reads $\mat S^{-}=[\mat S+\varepsilon\e{-\mat S/\varepsilon_S}]^{-1}$.
We shall name this method  ``{\Regtwo}''.

In both approaches {\Regone} and {\Regtwo}, $\varepsilon_S$ is chosen to be small enough so that the approximation $\mat S\approx[\mat S^{-}]^{-1}$ is valid, and chosen large enough so that $\mat S^{-}$ exists.

\subsection{Model}

The model we employ for testing the different methods is a one-dimensional double-well potential described by the Hamiltonian
\bea\label{eq:Ham}
\op H & = & \frac{1}{2}\op P^2 - \frac{1}{5} \op Q^2 + \frac{1}{4} \op Q^4,
\eea
where $\op Q$ and $\op P$ are the dimensionless position and momentum operators, respectively, and all quantities are given in atomic units.
It shows two minima at $x_\pm\approx\pm 0.63$ separated by a barrier height of $0.04$.
To solve the \gls{SE} numerically ``exactly'', we project the solution and the Hamiltonian onto a time-independent basis of $n_b$ functions $\{\ket{n};n=1,\dots,n_b\}$ build as the solution of the Harmonic oscillator $\op H_0=\frac{1}{2}(\op P^2 + \op Q^2)$, with $\op H_0\ket{n}=n\ket{n}$.
We chose the initial condition to be a Gaussian state placed at the minimum $x_{-}$
\bea
\ket{\Psi(0)} & = & \e{-\iu\op P x_{-}}\ket{n=0}.
\eea
We employed $n_b=20$ basis functions to generate numerically exact results.
To assess the convergence of the Gaussian-based methods, we compare the time evolution of two quantities.
The first quantity is the average position $\braOket{\Psi(t)}{\op Q}{\Psi(t)}$, which assesses the ability of the method to transfer the population from one well to the other.
The second quantity is the magnitude of the autocorrelation function $|\braket{\Psi(0)}{\Psi(t)}|$, which assesses the capacity of the time-dependent basis to cover the accessible eigenfunctions along the dynamics.
We will also study symmetry properties of the propagation: unitarity, time-reversibility, and energy conservation.

\subsection{Definition of the Gaussian basis}

We chose the Gaussian basis as coherent states of $\op H_0$ parameterized by complex numbers $\{z_k(t);k=1,\dots,N_g\}$ such that they satisfy the eigenequation
\bea
\op a \ket{g_k} & = & z_k \ket{g_k},
\eea
where $\op a=(\op Q+\iu\op P)/\sqrt{2}$ is the annihilation operator of the Harmonic oscillator described by $\op H_0$.
The Gaussian parameters are combinations of positions $q_k(t)$ and momenta $p_k(t)$, $z_k=(q_k+\iu p_k)/\sqrt{2}$, in the sense that
$\braOket{g_k}{\op Q}{g_k} = q_k$, and
$\braOket{g_k}{\op P}{g_k} = p_k$.

The time evolution of $q_k(t)$ and $p_k(t)$ is given by a classical \gls{EOM}, using the Hamilton function $H_k=p_k^2/2 - q_k^2/5 + q_k^4/4$, which can be expressed compactly as
\bea\label{eq:zdot}
\dot z_k & = & -\iu\left( z_k - \frac{7}{5}\Re[z_k] + \frac{1}{2}\Re[z_k]^3 \right).
\eea

The initial positions and momenta of all the Gaussians are generated using a Monte Carlo sampling algorithm:
the position is taken randomly in the interval $[-1:1]$ and the momentum is chosen as the inverse of the kinetic energy Boltzmann distribution.
The corresponding phase space configuration is accepted or rejected according to the thermal energy $k_B T=0.2$ a.u., where $k_B$ is the Boltzmann constant and $T$ the chosen temperature.

\subsection{Integration schemes}

The exact propagator appearing in \eq{eq:exactpropag}, $\aU_{21}$, requires evaluating matrix elements of the form
\bea
\braOket{g_k(t_2)}{\e{-\iu\op H(t_2-t_1)}}{g_l(t_1)},
\eea
which cannot be evaluated analytically for a general Hamiltonian.
We rather approximate the propagator as an expansion in $t_2-t_1$ in the spirit of the Crank-Nicolson integrator, similarly to 
Ref.~\citenum{Maskri:2022/ptrsa/20200379}.
We approximate \eq{eq:exactPsi2} as the symmetric form
\bea
[ \op 1 + \iu \op H \frac{t_2-t_1}{2} ] \ket{\Psi_2^e} & \approx & [ \op 1 - \iu \op H \frac{t_2-t_1}{2} ] \ket{\Psi_1},
\eea
which allows for a time-reversible integration.
We then project onto the basis at time $t_2$:
\bea
[ \mat 1_{M_2} + \iu \bar{\mat H}_2 \frac{t_2-t_1}{2} ] \bar{\mat C}_2 & \approx & [ \bar{\mat S}_{21} - \iu \bar{\mat H}_{21} \frac{t_2-t_1}{2} ] \bar{\mat C}_1, 
\eea
where $\bar{\mat S}_{21}=\mat\varphi_2^* \mat\varphi_1^t$ and $\bar{\mat H}_{21}=\mat\varphi_2^* \op H \mat\varphi_1^t$.
We could also have chosen to project on the basis at time $t_1$ to obtain
\bea
&[ \bar{\mat S}_{21}^\dagger + \iu \bar{\mat H}_{21}^\dagger \frac{t_2-t_1}{2} ] \bar{\mat C}_2 \approx [ \mat 1_{M_1} - \iu \bar{\mat H}_{1} \frac{t_2-t_1}{2} ] \bar{\mat C}_1. 
\eea
Thus, we have two comparable approximate propagators that we can use to approximate unitary evolution by minimizing the quantity
\bea
&\frac{1}{2}\norm{[ \mat 1_{M_2} + \iu \bar{\mat H}_2 \frac{t_2-t_1}{2} ] \bar{\mat U}_{21} - [ \bar{\mat S}_{21} - \iu \bar{\mat H}_{21} \frac{t_2-t_1}{2} ]}^2 + \nonumber\\
& \frac{1}{2}\norm{[ \bar{\mat S}_{21}^\dagger + \iu \bar{\mat H}_{21}^\dagger \frac{t_2-t_1}{2} ] \bar{\mat U}_{21} - [ \mat 1_{M_1} - \iu \bar{\mat H}_{1} \frac{t_2-t_1}{2} ]}^2. 
\eea
We can then apply the same procedure as defined in Sec.~\ref{sec:theory}, but with the following definition of $\aU_{21}$:
\bea\label{eq:Utilde12}
2 \aU_{21} 
& = & [ \mat 1_{M_2} - \iu \bar{\mat H}_2 \frac{t_2-t_1}{2} ] [ \bar{\mat S}_{21} - \iu \bar{\mat H}_{21} \frac{t_2-t_1}{2} ] + \nonumber\\
&& [ \bar{\mat S}_{21} - \iu \bar{\mat H}_{21} \frac{t_2-t_1}{2} ] [ \mat 1_{M_1} - \iu \bar{\mat H}_1 \frac{t_2-t_1}{2} ].
\eea

We chose a similar scheme to integrate the \gls{EOM} \eq{eq:Cdot}.
Combining forward and backward Euler steps, we obtain:
\bea
\mat C_2 & = & \left[\mat 1_{M_2} + \frac{t_2-t_1}{2} \mat S_2^{-} \left(\iu \mat H_2+\mat \tau_2\right)\right]^{-1} \nonumber\\
&&\times\left[\mat 1_{M_1} - \frac{t_2-t_1}{2} \mat S_1^{-} \left(\iu \mat H_1+\mat \tau_1\right)\right]\mat C_1.
\eea

For the classical evolution of the Gaussians, the \gls{EOM} given in \eq{eq:zdot} is also numerically solve using a Crank-Nicolson integration scheme.
Combining the forward and backward Euler steps applied on \eq{eq:zdot} results in the implicit scheme
\bea
\mat z_2 & = & \mat z_1 + \frac{t_2-t_1}{2} \left( \dot{\mat z}_1 + \dot{\mat z}_2 \right). \label{eq:implicit}
\eea
We solve this non-linear equation using the Newton method with the Euler forward step, $\mat z_2=\mat z_1 + (t_2-t_1) \dot{\mat z}_1$, as the initial guess to the solution.

All simulations are done with the Octave package~\cite{octave}.

\section{Results and discussion}
\label{sec:results}

To quantitatively analyze the performance of the new approach, we define an error function
\bea\label{eq:LocErr}
\Delta[f(\Psi[t])] & = & f(\Psi[t]) - f(\Psi_{exact}[t]),
\eea
of the quantity $f(\Psi[t])$ of interest, which depends on the time-dependent wave-function $\Psi[t]$.
In the following, we will consider the quantity $f(\Psi[t])$ to be: the autocorrelation function $\braket{\Psi[0]}{\Psi[t]}$, the energy $\braOket{\Psi[t]}{\op H}{\Psi[t]}$, the average position $\braOket{\Psi[t]}{\op Q}{\Psi[t]}$, or the overlap between two distinct states $\braket{\Psi[t]}{\Psi'[t]}$.
To analyze this error in a compact form for the entire dynamics, we define the time average error as
\bea\label{eq:AveErr}
\bar\Delta_t[f(\Psi)] & = & \frac{1}{t_f-t_i}\int_{t_i}^{t_f} dt\,|\Delta[f(\Psi[t])]|.
\eea

\subsection{Dynamics and convergence}

\begin{figure}
\includegraphics[width=0.5\textwidth]{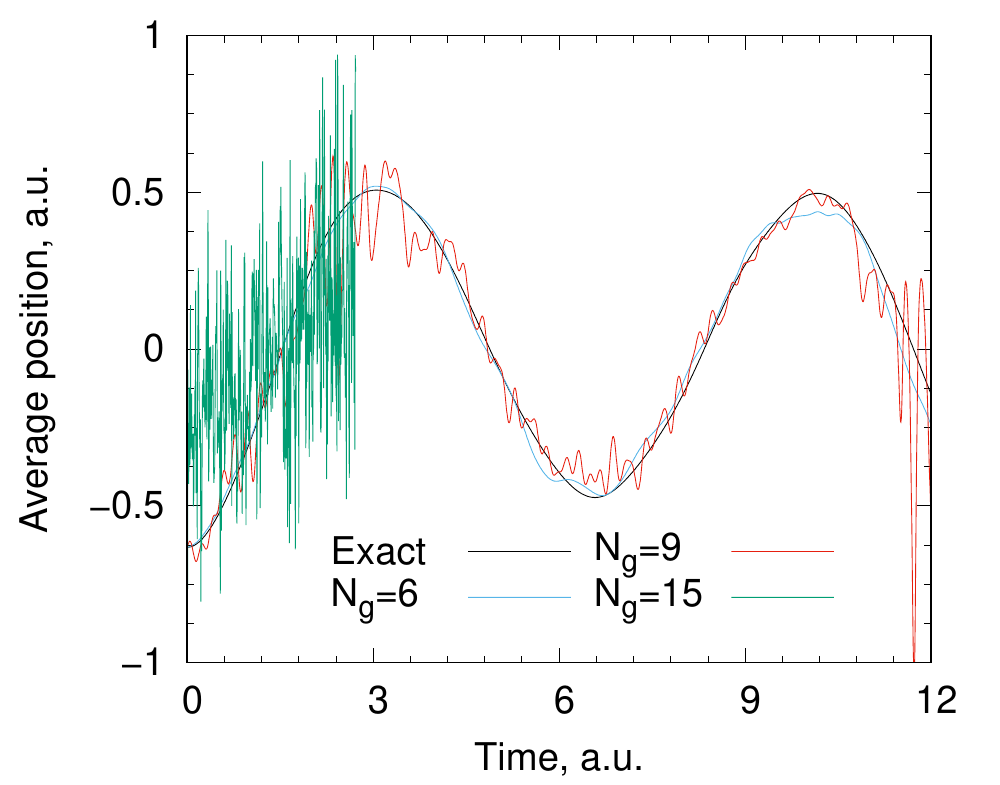}
\caption{Position expectation value using the naive method where the overlap inversion is attempted.}
\label{fig:position-orig}
\end{figure}

\begin{figure}
\includegraphics[width=0.5\textwidth]{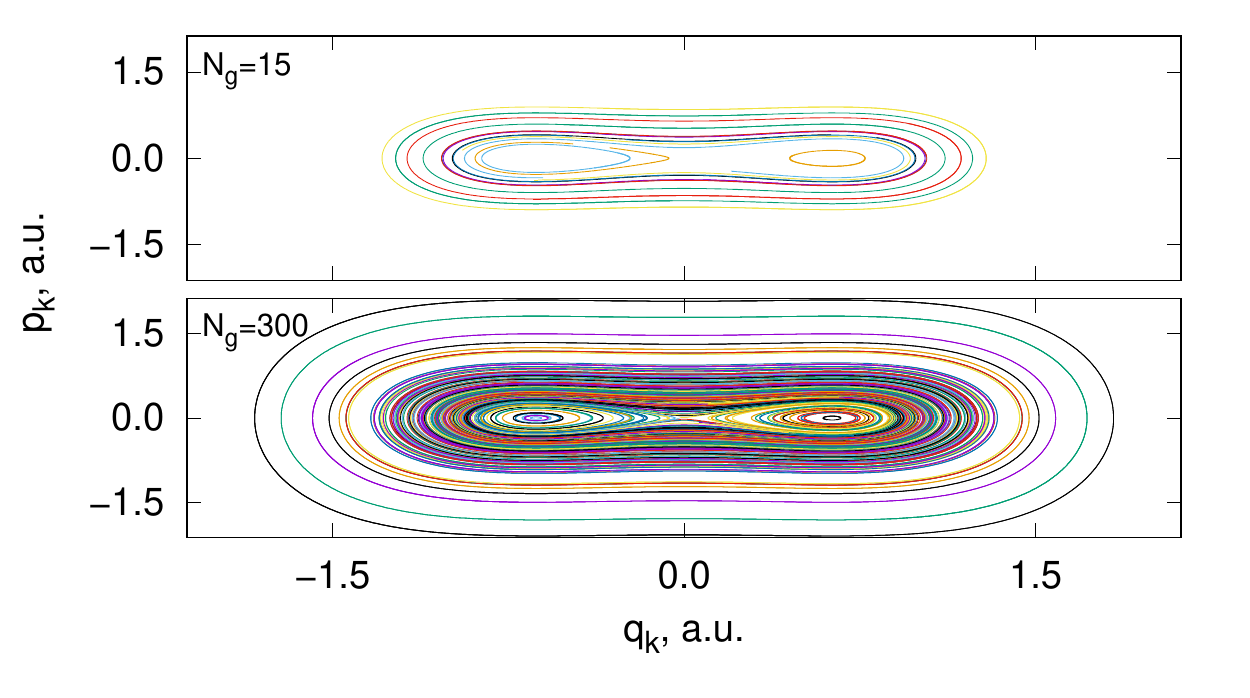}
\caption{Dynamics of the positions $q_k(t)$ and momenta $p_k(t)$ of the basis functions in the case $N_g=15$ and $N_g=300$.}
\label{fig:phasespacez}
\end{figure}

We first observe that the original approach given by \eq{eq:Cdot} fails.
We would expect that, since \eq{eq:Cdot} is derived from a variational principle, increasing the size of the basis should bring the resulting time evolution closer to the exact result.
Nevertheless, we do not observe convergence due to the singularity in the overlap matrix, which causes large numerical errors in solving the linear system \eq{eq:Cdot}.
The time evolution of the expectation value of the position, $\braOket{\Psi(t)}{\op Q}{\Psi(t)}$, is given in Fig.~\ref{fig:position-orig} for various number of Gaussian basis functions, $N_g$, and compared to the exact time-evolution.
It clearly shows how the dynamics generated by \eq{eq:Cdot} diverges from the exact one as the number of basis functions is increased.
We even observe that the dynamics with $N_g=15$ is stopped before $t=3$ a.u.
The root of this divergence is numerical instabilities in inverting the overlap matrix when near-linear dependencies appear in the Gaussian basis set.
Indeed, 2 eigenvalues of the overlap matrix are lower than $10^{-7}$ when $N_g=9$ and this number increases to 7 when $N_g=15$.
The phase-space trajectories of the $N_g=15$ parameters $z_k$ are represented in Fig.~\ref{fig:phasespacez}.
It clearly shows that 15 Gaussians are not sufficient for covering the accessible phase-space such that the dynamics with $N_g=15$ Gaussian basis functions is far from being converged.

\begin{figure}
\includegraphics[width=0.5\textwidth]{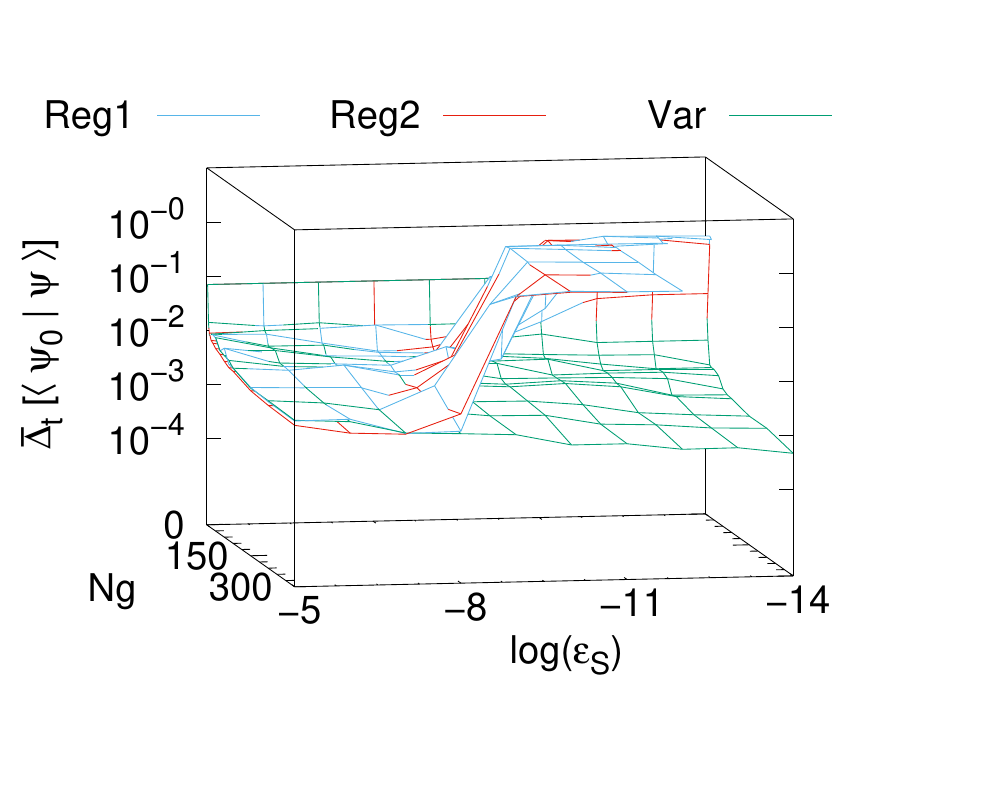}
\caption{Time average errors on the autocorrelation function $\bar{\Delta}_t[\braket{\Psi_0}{\Psi}]$ (see \eq{eq:AveErr}) using the regularized and the variational methods for different number of Gaussian basis functions $N_g$ and different value of $\varepsilon_S$.
The plateau visible for small values of $\varepsilon_s$ for the regularized methods depicts the divergences of the errors.}
\label{fig:autocorrel}
\end{figure}

Upon Fourier transform, the autocorrelation function is directly related to the eigenspectrum of the Hamiltonian that is accessible by a given initial state.
Hence, the ability of a simulation to reproduce the exact time evolution of the autocorrelation function is directly related to the level of convergence.
We ran a systematic study of convergence by evaluating the time-average error on the autocorrelation function's modulus, $f(\Psi[t])=|\braket{\Psi(0)}{\Psi(t)}|$, for $N_g\in[3:300]$. 
The ability of the method to converge to the exact solution is impacted by the value of the threshold $\varepsilon_S$ used in regularizing the overlap matrix or to define the linearly independent set of basis functions.
Figure~\ref{fig:autocorrel} gives the time-average error on the autocorrelation function's modulus, $f(\Psi[t])=|\braket{\Psi(0)}{\Psi(t)}|$, with respect to $N_g$ and $\varepsilon_S$.
This figure clearly shows that regularized approaches become unstable (diverges) for $\varepsilon_S<10^{-8}$ while the variational one remains stable.
Thanks to its stability, the variational approach utilizes a linearly independent subspace where basis functions are distinguished down to the numerical error of the method $~10^{-14}$, which allows for an optimal convergence.
This extra stability of the variational approach is possible because the inverse of the overlap matrix does not appear in the equation and is replaced by the inverse of its square root.
As a result, the variational approach converges at a lower error than the regularized approaches for the same $N_g$.

\begin{figure}
\includegraphics[width=0.5\textwidth]{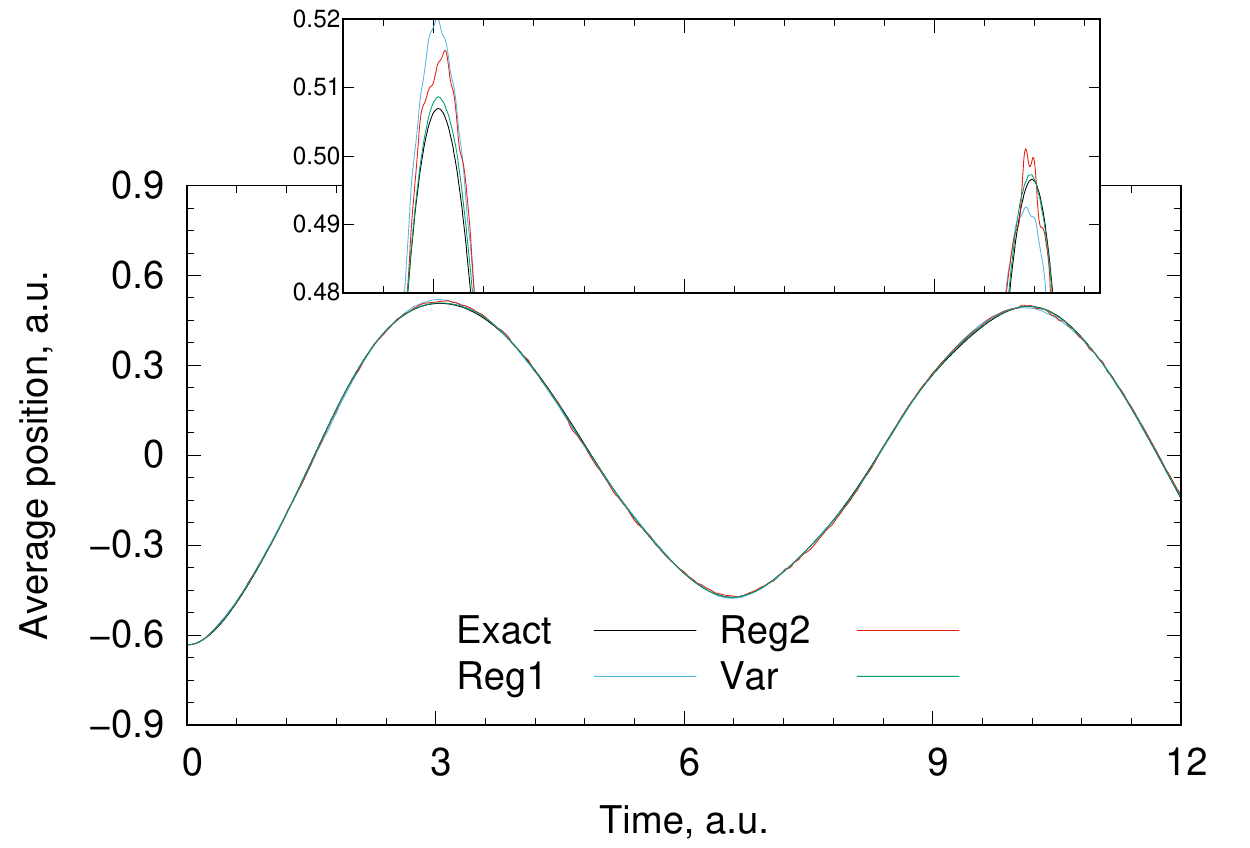}
\caption{Position expectation value for $N_g=300$ using the regularized methods with $\varepsilon_S=10^{-7}$ and the variational method with $\varepsilon_S=10^{-14}$. The upper frame is a zoom on the details of the curves at the maxima.}
\label{fig:position-others}
\end{figure}

The best case scenarios for each of the methods are $N_g=300$ with $\varepsilon_S=10^{-7}$ for the regularized methods ({\Regone} and {\Regtwo}), and with $\varepsilon_S=10^{-14}$ for the {\Var} method.
The time evolution of the average position for these 3 cases is given in Fig.~\ref{fig:position-others} and shows that all 3 methods are qualitatively correct.
By zooming on the curve, for example at the maxima as done in Fig.~\ref{fig:position-others}, we can see that the regularized methods exhibit small discrepancies as opposed to the variational method that is quantitatively correct.
This is in line with the observation made in the previous paragraph.

\subsection{Conserved quantities}

It is well known that the exact solution of the \gls{SE} conserves various quantities under specific conditions (e.g. energy conservation and unitarity for a closed system).
In this section, we investigate the ability of the different methods to conserve the following quantities: i) the scalar product (unitarity), ii) the mean energy, and iii) the dynamics under time inversion (time-reversibility).

\subsubsection{Unitarity}

\begin{figure}
\includegraphics[width=0.5\textwidth]{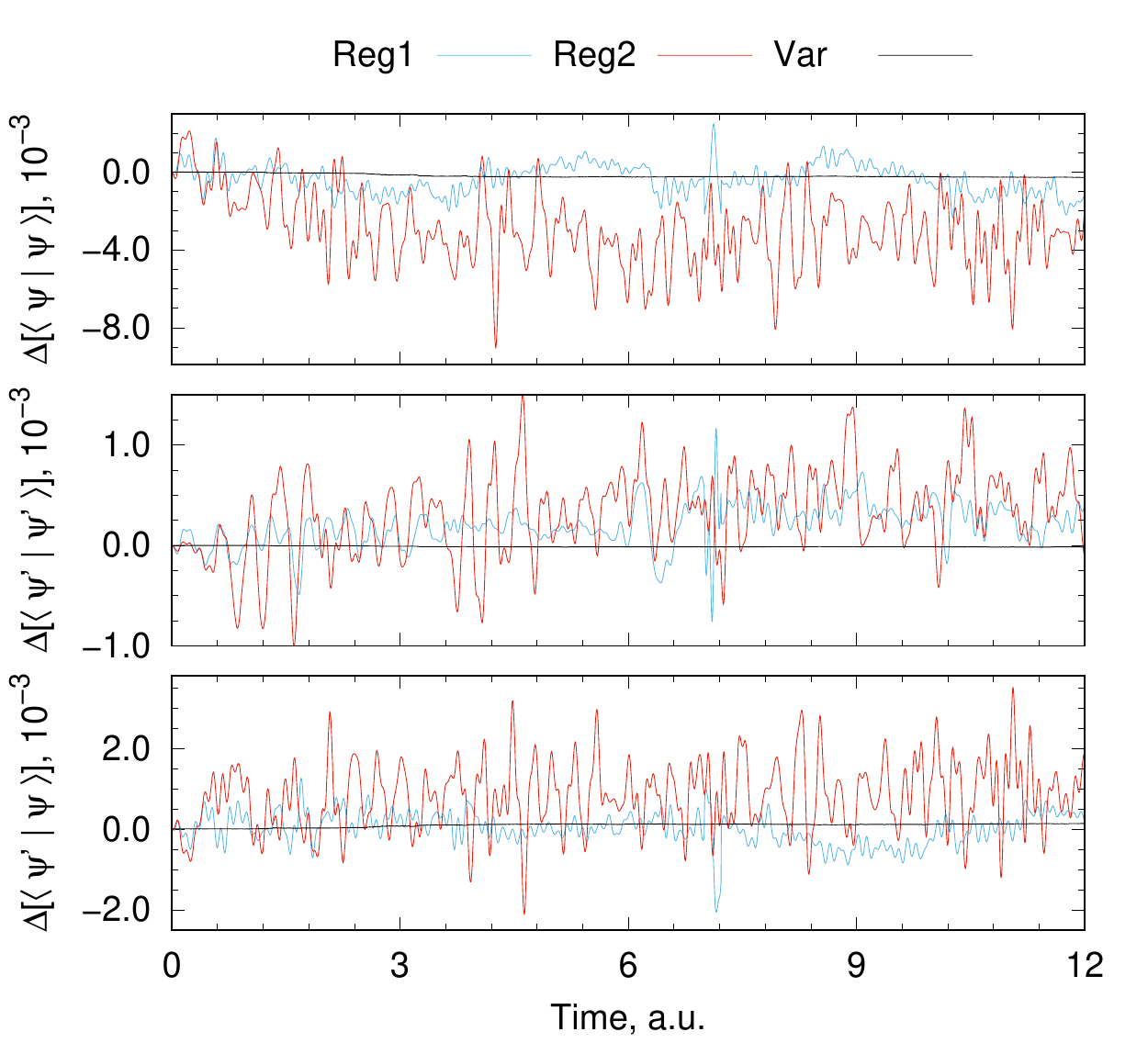}
\caption{Time evolution of the errors (see \eq{eq:LocErr}) on norms of $\ket{\Psi}$ and $\ket{\Psi'}$, and their scalar product for $N_g=300$ using the regularized methods with $\varepsilon_S=10^{-7}$ and the variational method with $\varepsilon_S=10^{-14}$.}
\label{fig:unitarity}
\end{figure}

To test unitarity of the method, we define a second wavefunction $\ket{\Psi'}$ and study the time evolution of the norms $\norm{\ket{\Psi}}$ and $\norm{\ket{\Psi'}}$, and the normalized scalar product $|\braket{\Psi}{\Psi'}|/(\norm{\ket{\Psi}}\norm{\ket{\Psi'}})$.
This second wavefunction is defined by its initial condition where $z_2=-0.034575-\iu 0.521422$
\bea
\ket{\Psi'(0)} & = & \e{z_2\op a^\dagger-z_2^*\op a}\ket{n=0}.
\eea
The results, given in Fig.~\ref{fig:unitarity}, show that the variational method conserves these quantities with an error of $10^{-4}$, while the regularized methods do not conserve these quantities with an error of at least $10^{-3}$.
The variational method designed in the current document is in fact built to be strictly unitary in the subspace of linearly independent basis functions $\{\ket{\varphi(t)}\}$.
Nevertheless, to propagate the coefficients vector $\mat C(t)$, it is necessary to multiply the unitary transformation $\bar{\mat U}(t_2,t_1)$ [given by \eq{eq:U12bar}] by the inverse of the overlap square root on the left. This operation considerably alters the numerical stability if $\varepsilon_S$ is very small, as is the case in Fig.~\ref{fig:unitarity} where $\varepsilon_S=10^{-14}$ for the variational method.
In fact, we can make the error on the scalar product arbitrarily small, down to $10^{-12}$, if we increase this threshold to $\varepsilon_S=10^{-7}$.
We can also observe that the method {\Regone} usually performs better than {\Regtwo} with a smaller amplitude of variations.

\subsubsection{Energy conservation}

\begin{figure}
\begin{picture}(100,200)
\put(-75,0){\includegraphics[width=0.5\textwidth]{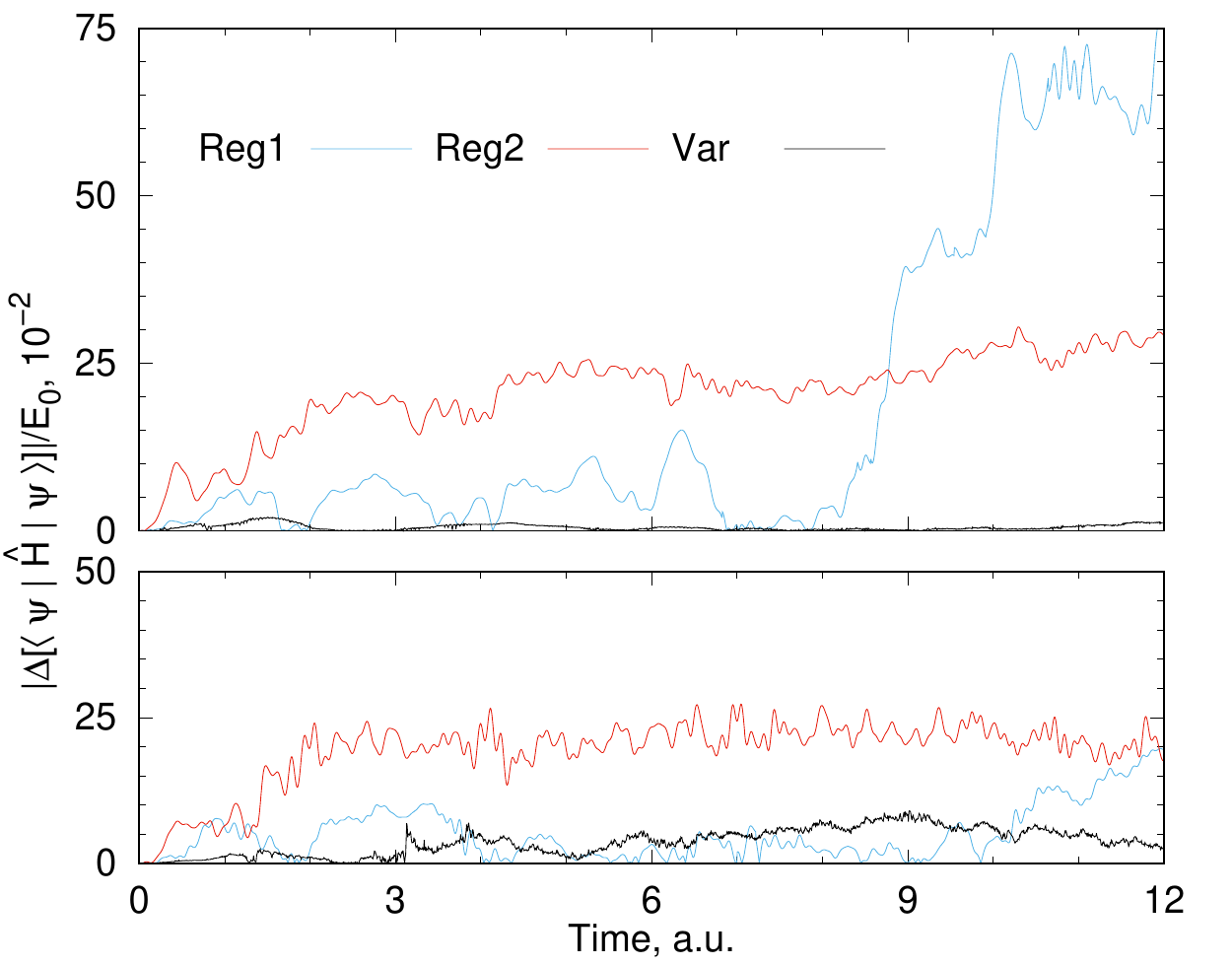}}
\put(-40,185){a)}
\put(-40,72){b)}
\end{picture}
\caption{Time evolution of the relative error on the energy (see \eq{eq:LocErr}) for $N_g=300$ using: 
a) $\varepsilon_S=10^{-6}$ and the variational method with $\varepsilon_S=10^{-13}$ for the regularized methods, and
b) $\varepsilon_S=10^{-7}$ and the variational method with $\varepsilon_S=10^{-14}$ for the regularized methods.
Here, $E_0=0.0975$ a.u. is the initial energy.}
\label{fig:energy}
\end{figure}

It can be shown that in approaches where the time-dependent basis evolves variationally, the obtained \gls{EOM} for closed systems conserves energy.~\cite{Kramer:1981/Book,Beck:2000/pr/1,Habershon:2012/jcp/014109,Joubert:2015/jcp/134107,Hackl:2020/sp/048}
However, this is not the case in the current document and energy is not, a priori, conserved.
It can be shown that the error on energy conservation should diminish as the basis approaches completeness.
In general, increasing the number of Gaussians introduces a strong linear dependence, which causes numerical instabilities in \eq{eq:EOMSC}, and reaching energy conservation seems hopeless.
Nevertheless, the new approach does not suffer from this difficulty and should show a better conservation of energy as the size of the linearly independent space increases.
On the contrary, the regularized methods are not variational and energy conservation is not ensured when the time-dependent basis reaches convergence.
This is what we observed in Fig.~\ref{fig:energy} where the variational method clearly shows less deviation of the energy along the dynamics.

While the differences are not quantitative in the figure, we must emphasize that, even if the method {\Var} is variational, the reduction of the error on the energy \emph{is not variational} as it can be seen in Tab.~{\ref{tab:energy}.
In particular we can observe that the calculation with $N_g=300$ and $\varepsilon_S=10^{-13}$ gives a much lower error for the method {\Var}, which is an order of magnitude lower than the error of the simulation in Fig.~\ref{fig:energy}-a) compared to the case where $\varepsilon_S=10^{-14}$ in Fig.~\ref{fig:energy}-b).
As a comparison, results from the other methods are also given in Tab.~{\ref{tab:energy}, and show that errors using {\Regone} and {\Regtwo} are generally larger by an order of magnitude (or more) compared to the {\Var} method.

\begin{table}
\caption{Table of the time-average energy errors, $\bar\Delta_t[\braket{\Psi}{\op H\Psi}]$, using different values of $N_g$ and $\varepsilon_S$ for the three methods.
Results are given in $10^{-3}$ a.u.}
\label{tab:energy}
\setlength{\tabcolsep}{0.27cm}
\begin{tabular}{|l|c|c|c|c|}
\hline
Method  & $\varepsilon_S$ & $N_g=150$ & $N_g=210$ & $N_g=300$ \\\hline
        &   $10^{-12}$    &  0.47     &  1.45     &  1.80     \\
  \Var  &   $10^{-13}$    &  4.15     &  0.88     &  0.51     \\
        &   $10^{-14}$    &  2.71     &  0.72     &  3.59     \\\hline
        &   $10^{-5}$     & 19.38     &  7.42     &  7.65     \\
\Regone &   $10^{-6}$     &  3.44     &  6.66     & 18.22     \\
        &   $10^{-7}$     & 14.99     & 16.47     &  4.92     \\\hline
        &   $10^{-5}$     & 27.84     & 30.67     & 20.85     \\
\Regtwo &   $10^{-6}$     & 18.65     & 22.13     & 20.04     \\
        &   $10^{-7}$     & 25.06     & 17.54     & 18.79     \\\hline
\end{tabular}
\end{table}

\subsubsection{Time-reversibility}

The original \gls{EOM} is infinitesimally time-reversible and our choice of integrator transfers this time-reversibility for a finite time step.
This time-reversibility is again conserved by construction for the regularized method.
On the contrary, the variational method looses this time-reversibility because the size of the linearly independent subspace varies over time.
Time-reversible methods are known to be more stable and we want to evaluate how the {\Var} method performs.
To this aim, we propagated the wavefunction obtained at time $t=6$ a.u. backward in time back to $t=0$ a.u.
The comparison between the forward and the backward propagation is made based on the time evolution of the average position.
Figure~\ref{fig:trevert-error} presents the errors between the forward and the backward time evolution for the different methods.
It confirms the time-reversibility of the methods {\Regone} and \Regtwo, but it also shows that the error on the time-reversibility for the method {\Var} is rather low and at most $2\cdot 10^{-4}$ a.u.
In fact, the time-dependent position expectation value obtained from the backward dynamics is visually indistinguishable from the forward, as shown in Figure~\ref{fig:trevert}.
\begin{figure}
\includegraphics[width=0.5\textwidth]{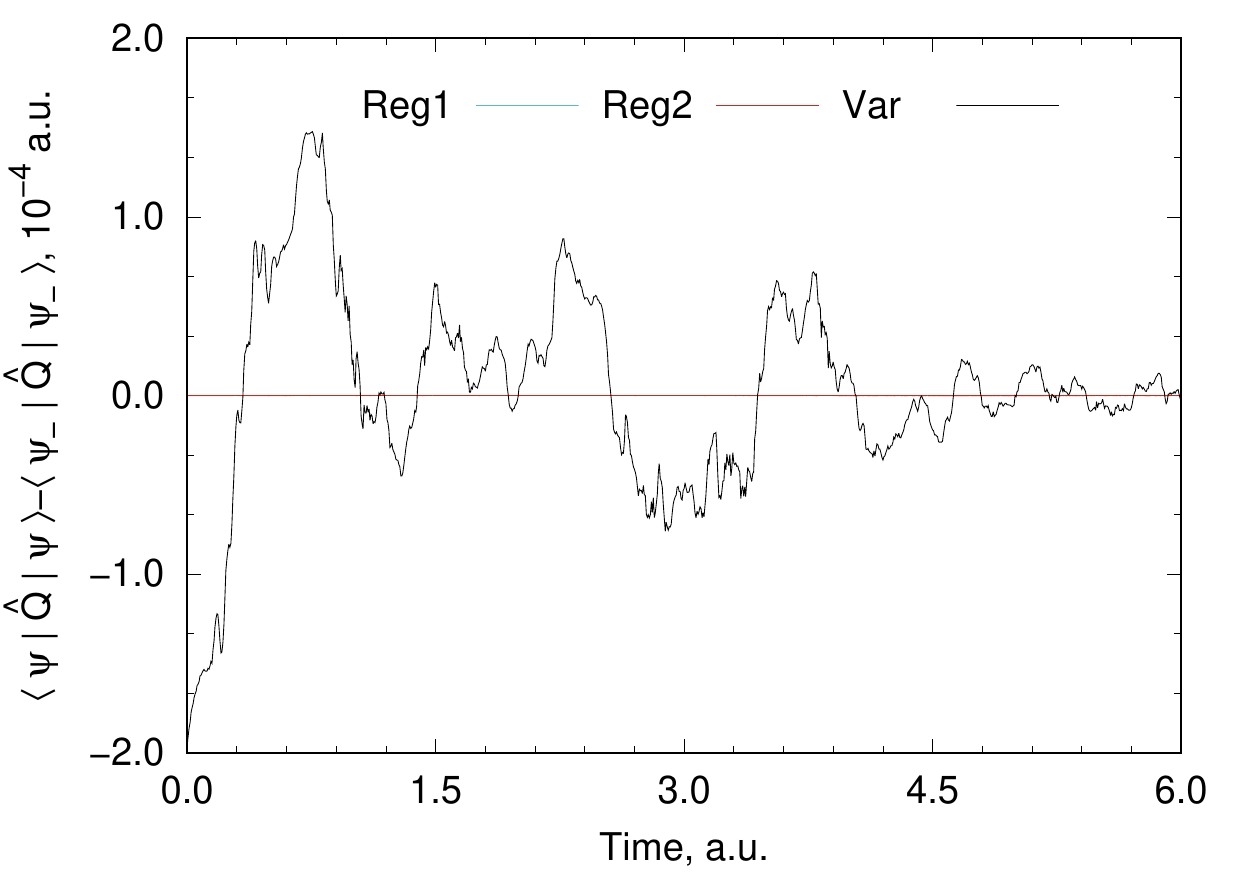}
\caption{Difference in average position evaluated during forward and backward in time propagations, for $N_g=300$ using the regularized methods with $\varepsilon_S=10^{-7}$ and the variational methods with $\varepsilon_S=10^{-14}$.}
\label{fig:trevert-error}
\end{figure}
\begin{figure}
\includegraphics[width=0.5\textwidth]{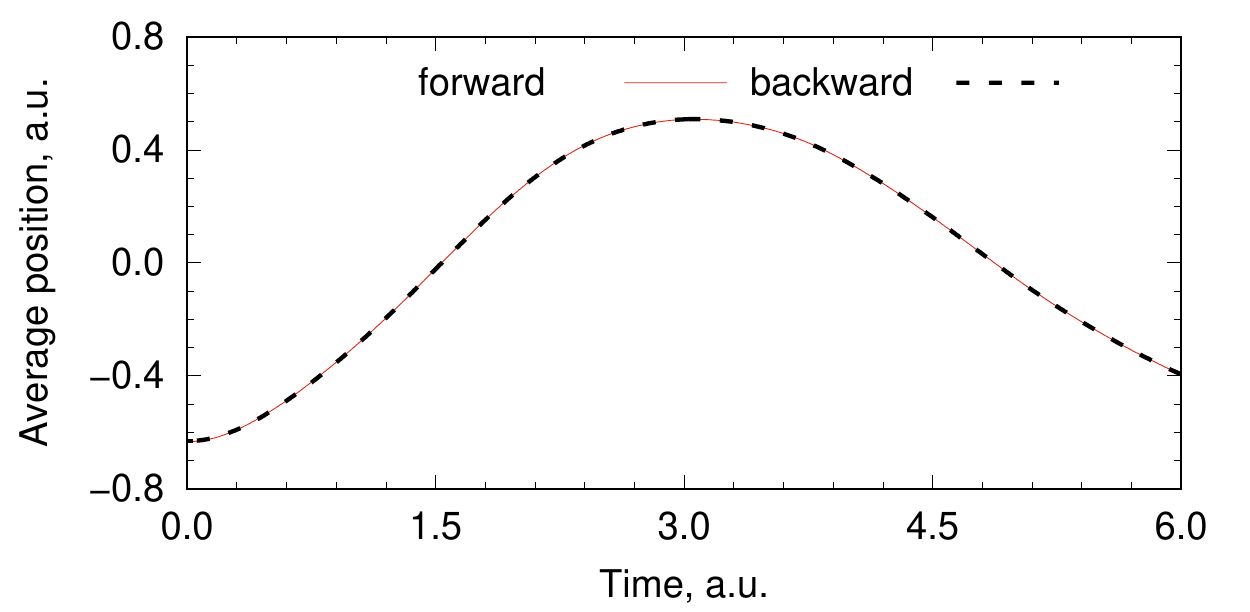}
\caption{Average position evaluated during the forward and backward in time propagations using the variational method for $N_g=300$ with $\varepsilon_S=10^{-14}$.}
\label{fig:trevert}
\end{figure}

\subsection{Convergence properties for a model with a reduced linear dependence}

Numerical simulations are generally done with basis functions that are optimized to diminished linear dependence.
To test the {\Var} method in a more realistic context, we modified the Hamiltonian to decrease the linear dependence and investigate the convergence in this case.
For this purpose, we artificially rescaled the position operator, while we still employ coherent states of $\op H_0=(\op P^2+\op Q^2)/2$.
The model Hamiltonian becomes
\bea\label{eq:Hp}
\op H' & = & \frac{1}{2}\op P^2 - \frac{1}{8} \op Q^2 + \frac{1}{8^2} \op Q^4,
\eea
where minima are now $x'_\pm=\pm 2$ a.u.
Hence, the rescale factor is about $3$ compared to \eq{eq:Ham}.
All parameters and simulation details are otherwise the same as already given in Sec.~\ref{sec:details}.
Time-average errors (see definitions in Eq.~(\ref{eq:AveErr})) from a $31.739$ a.u. time propagation with this new Hamiltonian are given in Tab.~\ref{tab:results2}.
They show that numerical precision is almost met regarding norm and energy conservation for the {\Var} method, while the regularized methods still show significant deviations.
Regarding the correlation function and the average position, the {\Var} method shows an error that is an order of magnitude smaller than the {\Regone} and {\Regtwo} methods.
\begin{table}
\caption{Time-average errors, $\bar\Delta_t[f(\Psi)]$ (see \eq{eq:AveErr}) given in atomic units, for various quantities described by the functions $f$, with $N_g=300$, and $\varepsilon_S=10^{-14}$ for {\Var} and $\varepsilon_S=10^{-7}$ for {\Regone} and \Regtwo, and using the Hamiltonian in \eq{eq:Hp}.}
\label{tab:results2}
\setlength{\tabcolsep}{0.1cm}
\begin{tabular}{|l|c|c|c|c|}
\hline
 method  &$f=\norm{\ket{\Psi}}$&$f=\braOket{\Psi}{\op H'}{\Psi}$&$f=\braket{\Psi_0}{\Psi}$&$f=\braOket{\Psi}{\op Q}{\Psi}$\\\hline
   \Var  & $7\cdot 10^{-12}$ & $3\cdot 10^{-9}$ & $4\cdot 10^{-7}$ & $1\cdot 10^{-6}$ \\\hline
 \Regone & $3\cdot 10^{-6}$  & $6\cdot 10^{-7}$ & $2\cdot 10^{-6}$ & $1\cdot 10^{-5}$ \\\hline
 \Regtwo & $4\cdot 10^{-6}$  & $1\cdot 10^{-6}$ & $2\cdot 10^{-6}$ & $1\cdot 10^{-5}$ \\\hline
\end{tabular}
\end{table}

\section{Conclusion}
\label{sec:conclusion}

In this paper, we show that the problem of linear dependence can be handled by defining a time-dependent working space with varying dimensionality.
Introducing this change in dimensionality implies that evolution cannot be unitary and time-reversible at the same time.
Instead, we define a variational approach over finite time steps, that results into a projection of the time evolution transformation onto the Stiefel manifold of semi-unitary transformations.

The resulting method variationally evolves the solution of the \gls{SE} using the maximum information from the linearly dependent \glspl{TDBF}.
Furthermore, the evolution is unitary and time-reversible when the dimensionality of the working space is constant.
When the dimensionality varies, the evolution becomes semi-unitary and is not time-reversible anymore.
The time-reversibility for the method could be recovered by imposing a fixed size for the linearly independent subspace.
This is essentially what the 2-layer \gls{GMCTDH} method~\cite{Romer:2013/jcp/064106,Richings:2015/irpc/269,Eisenbrandt:2018/jcp/174101} does if a ``linear'' parameterization is employed.
However, we generally do not know this fixed size in advance and further investigations are required to estimate a size that is sufficiently large to help convergence and sufficiently low to avoid numerical instabilities.
In this sense, the variational method presented in this paper takes full advantage of the \glspl{TDBF} by allowing the dimensionality to change.

While the method is not time-reversible and does not conserve energy, we show that errors on these two aspects can be diminished thanks to the fact that large basis set can be employed to help reach convergence.
Furthermore, the (semi-)unitarity of the method ensures that the norm is conserved and the method is more stable.
Although this method was developed to handle \glspl{TDBF} exhibiting strong linear dependence, it also shows better performance compared to methods employing regularization when linear dependence is less severe.
Regarding convergence, one advantage of the new method is that only inversion of the square-root of the overlap matrix is needed (instead of the inverse of the overlap matrix itself), which improves the condition number in the inversion.

Finally, we note that the presented methodology is rather general and could be combined with the usual \gls{TDVP} to avoid difficulties in similar linear dependencies appearing, for example, in the \gls{MCTDH} method.

\section{Acknowledgments}

The author is grateful to Caroline Lasser for stimulating discussions, and to \'Etienne Mangaud, Kossi Kety, and Mina Asaad for their critical comments on the manuscript.
The author acknowledges support from the CNRS 80$|$Prime project AlgDynQua, the visiting professorship program of the I-Site Future, and financial support from French National Research Agency through the project ANR-21-CE29-0005.

\appendix

\section{Derivation of the \gls{EOM} in the linearly independent basis}
\label{app:proofCbdot}

Using Eqs.~(\ref{eq:Cb}-\ref{eq:Phim}) and \eq{eq:Phi}, we can rewrite that $\bar{\mat C}=\mat\Phi^\dagger\mat S\mat C$.
Hence, the time-derivative is given by
\bea\label{eq:deriv1}
\dot{\bar{\mat C}} 
&=&
  \dot{\mat\Phi}^\dagger\mat S\mat C
+ \mat\Phi^\dagger\dot{\mat S}\mat C
- \mat\Phi^\dagger\left[\mat\tau+\iu\mat H\right]\mat C,
\eea
where we utilized \eq{eq:EOMSC} to substitute $\mat S\dot{\mat C}$.
Then, using \eq{eq:Psinewform} and \eq{eq:Cb}, we have that $\mat C=\mat V\mat V^\dagger\mat C=\mat\Phi\bar{\mat C}$, so that \eq{eq:deriv1} becomes
\bea\label{eq:deriv2}
\dot{\bar{\mat C}} 
&=&
-\left[
  \mat\Phi^\dagger\mat\tau\mat\Phi 
- \dot{\mat\Phi}^\dagger\mat S\mat\Phi
- \mat\Phi^\dagger\dot{\mat S}\mat\Phi
+ \iu\bar{\mat H}
\right]\bar{\mat C}.
\eea
Finally, we use the fact that $\mat\varphi^*\mat\varphi^t=\mat\Phi^\dagger\mat S\mat\Phi=\mat 1_M$, which implies that $\dot{\mat\Phi}^\dagger\mat S\mat\Phi+\mat\Phi^\dagger\dot{\mat S}\mat\Phi+\mat\Phi^\dagger\mat S\dot{\mat\Phi}=\mat 0$, to obtain the form given in the main text \eq{eq:Cbdot}.

\section{Variational character of the method}
\label{app:variational}

Starting from \eq{eq:Lag1eq2}, we can express the trace using the definition of the projectors in \eq{eq:Pphi} at times $t_1$ and $t_2$ 
\bea
\tr\{\op P_1 (\op{\bar U}(t_2,t_1)-\op U(t_2,t_1))^\dagger \op P_2 (\op{\bar U}(t_2,t_1)-\op U(t_2,t_1))\}, \nonumber\\
\eea
where the operator $\op{\bar U}(t_2,t_1)$ is defined through the relation
\bea
\bar{\mat U}_{21} &=& \mat\varphi_2^*\op{\bar U}(t_2,t_1)\mat\varphi_1^t.
\eea
Taking the complete bases limits, we have $\op P_1\to 1$ and $\op P_2\to 1$, and the trace in \eq{eq:Lag1eq2} becomes
\bea\label{eq:normcomplete}
&\tr\{(\op{\bar U}(t_2,t_1)-\op U(t_2,t_1))^\dagger (\op{\bar U}(t_2,t_1)-\op U(t_2,t_1))\} = \nonumber\\&
\norm{\op{\bar U}(t_2,t_1)-\op U(t_2,t_1)}^2.
\eea
Since $\op U(t_2,t_1)$ is unitary, minimization of the norm in \eq{eq:normcomplete}, under the constraint that $\op{\bar U}(t_2,t_1)$ is unitary, is trivially achieved by the equality $\op{\bar U}(t_2,t_1)=\op U(t_2,t_1)$.
This equality shows that the minimization tends to the exact solution in the complete basis limit.

\section{Proof of equivalence with \gls{TDVP} for infinitesimal step}
\label{app:equiv}

In this appendix, we show how the new approach given by \eq{eq:properC2} is equivalent to the usual approach \eq{eq:Cbdot} in the limit of an infinitesimal time step and when $M_1=M_2=N_g$.
This is achieved by taking the time-derivative of \eq{eq:properC2} with respect to $t_2$ and comparing the result at $t_2=t_1=t$ with \eq{eq:Cdot}.
For simplicity we will work in the orthonormal basis $\{\varphi_k\}$ and use the form \eq{eq:Cbdot}.
Our starting point is the stationary condition that can be obtained for the Procrustes problem in the case $M_2=M_1$ where it is possible to show, by differentiating the error with respect to $\bar{\mat U}_{21}$, that the minimal error is achieved when the following equation is satisfied for $\bar{\mat U}_{21}$:
\bea\label{eq:statcond}
{\aU_{21}} - \bar{\mat U}_{21} {\aU_{21}}^\dagger \bar{\mat U}_{21} & = & \mat 0.
\eea
The time derivative $\dot{\bar{\mat U}}_{21}$ is obtained from the stationary condition given by differentiating \eq{eq:statcond}, which results in
\bea
\mat 0 
& = & \frac{\partial}{\partial t_2} \left[ {\aU_{21}} - \bar{\mat U}_{21} {\aU_{21}}^\dagger \bar{\mat U}_{21} \right] \nonumber\\
& = & \dot{\aU}_{21} 
- \bar{\mat U}_{21} ({\dot{\aU}_{21}})^\dagger \bar{\mat U}_{21}
- \dot{\bar{\mat U}}_{21} {\aU_{21}}^\dagger \bar{\mat U}_{21}
- \bar{\mat U}_{21} {\aU_{21}}^\dagger \dot{\bar{\mat U}}_{21}. \nonumber\\\label{eq:Dstatcond}
\eea
Using the definition of $\aU_{21}$ in \eq{eq:exactpropag} for $\op U(t_2,t_1)=\e{-\iu\op H (t_2-t_1)}$, we can evaluate $\dot{\aU}_{21}$
\bea
\dot{\aU}_{21} = \dot{\mat\varphi}_2^*\e{-\iu\op H (t_2-t_1)}\mat\varphi_1^t -\iu \mat\varphi_2^*\op H\e{-\iu\op H (t_2-t_1)}\mat\varphi_1^t.
\eea
Then, we take the limit $t_2\to t_1\equiv t$ for which $\mat\varphi_2\to \mat\varphi_1\equiv \mat\varphi$ and obtain 
\bea
\dot{\aU} = \bar{\mat\tau}^\dagger -\iu \bar{\mat H}.
\eea
Furthermore, in the same limit, we have that ${\bar{\mat U}}_{21}={\aU}_{21}=\mat 1_M$ so that \eq{eq:Dstatcond} becomes
\bea
\dot{\bar{\mat U}}
= \frac{1}{2} \left[ \dot{\aU} - ({\dot{\aU}})^\dagger \right] 
= \frac{1}{2} \left[ \bar{\mat\tau}^\dagger - \bar{\mat\tau} - 2 \iu \bar{\mat H} \right]
= - \bar{\mat\tau} - \iu \bar{\mat H}. \nonumber\\
\eea
In the last equality, we used the fact that $\mat\varphi^*\mat\varphi^t=\mat 1_M$ and thus $\bar{\mat\tau}^\dagger=-\bar{\mat\tau}$.
We are now in a position to evaluate $\dot{\bar{\mat C}}$ as follows
\bea
\dot{\bar{\mat C}} \equiv \frac{\partial\bar{\mat C}_2}{\partial t_2}\bigg|_{t_2=t} = \frac{\partial\bar{\mat U}_{21}}{\partial t_2} \bigg|_{t_2=t_1=t} \bar{\mat C} =  - [ \bar{\mat\tau} + \iu \bar{\mat H} ] \bar{\mat C}. \nonumber\\
\eea
In this last equality we recover \eq{eq:Cbdot} as expected.

\bibliography{Nonbijective}

\begin{thebibliography}{58}%
\makeatletter
\providecommand \@ifxundefined [1]{%
 \@ifx{#1\undefined}
}%
\providecommand \@ifnum [1]{%
 \ifnum #1\expandafter \@firstoftwo
 \else \expandafter \@secondoftwo
 \fi
}%
\providecommand \@ifx [1]{%
 \ifx #1\expandafter \@firstoftwo
 \else \expandafter \@secondoftwo
 \fi
}%
\providecommand \natexlab [1]{#1}%
\providecommand \enquote  [1]{``#1''}%
\providecommand \bibnamefont  [1]{#1}%
\providecommand \bibfnamefont [1]{#1}%
\providecommand \citenamefont [1]{#1}%
\providecommand \href@noop [0]{\@secondoftwo}%
\providecommand \href [0]{\begingroup \@sanitize@url \@href}%
\providecommand \@href[1]{\@@startlink{#1}\@@href}%
\providecommand \@@href[1]{\endgroup#1\@@endlink}%
\providecommand \@sanitize@url [0]{\catcode `\\12\catcode `\$12\catcode
  `\&12\catcode `\#12\catcode `\^12\catcode `\_12\catcode `\%12\relax}%
\providecommand \@@startlink[1]{}%
\providecommand \@@endlink[0]{}%
\providecommand \url  [0]{\begingroup\@sanitize@url \@url }%
\providecommand \@url [1]{\endgroup\@href {#1}{\urlprefix }}%
\providecommand \urlprefix  [0]{URL }%
\providecommand \Eprint [0]{\href }%
\providecommand \doibase [0]{http://dx.doi.org/}%
\providecommand \selectlanguage [0]{\@gobble}%
\providecommand \bibinfo  [0]{\@secondoftwo}%
\providecommand \bibfield  [0]{\@secondoftwo}%
\providecommand \translation [1]{[#1]}%
\providecommand \BibitemOpen [0]{}%
\providecommand \bibitemStop [0]{}%
\providecommand \bibitemNoStop [0]{.\EOS\space}%
\providecommand \EOS [0]{\spacefactor3000\relax}%
\providecommand \BibitemShut  [1]{\csname bibitem#1\endcsname}%
\let\auto@bib@innerbib\@empty
\bibitem [{\citenamefont {Frenkel}(1934)}]{Frenkel:1934/Book}%
  \BibitemOpen
  \bibfield  {author} {\bibinfo {author} {\bibfnamefont {J.}~\bibnamefont
  {Frenkel}},\ }\href@noop {} {\emph {\bibinfo {title} {Wave Mechanics}}}\
  (\bibinfo  {publisher} {Clarendon Press},\ \bibinfo {address} {Oxford},\
  \bibinfo {year} {1934})\BibitemShut {NoStop}%
\bibitem [{\citenamefont {Dirac}(1958)}]{Dirac:1958/Book}%
  \BibitemOpen
  \bibfield  {author} {\bibinfo {author} {\bibfnamefont {P.~A.~M.}\
  \bibnamefont {Dirac}},\ }\href@noop {} {\emph {\bibinfo {title} {The
  Principles of Quantum Mechanics, 4th Edition}}}\ (\bibinfo  {publisher}
  {Clarendon Press},\ \bibinfo {address} {Oxford},\ \bibinfo {year}
  {1958})\BibitemShut {NoStop}%
\bibitem [{\citenamefont {{McLachlan}}(1964)}]{Mclachlan:1964/mp/39}%
  \BibitemOpen
  \bibfield  {author} {\bibinfo {author} {\bibfnamefont {A.}~\bibnamefont
  {{McLachlan}}},\ }\href {\doibase 10.1080/00268976400100041} {\bibfield
  {journal} {\bibinfo  {journal} {Mol. Phys.}\ }\textbf {\bibinfo {volume}
  {8}},\ \bibinfo {pages} {39} (\bibinfo {year} {1964})}\BibitemShut {NoStop}%
\bibitem [{\citenamefont {Kramer}\ and\ \citenamefont
  {Saraceno}(1981)}]{Kramer:1981/Book}%
  \BibitemOpen
  \bibfield  {author} {\bibinfo {author} {\bibfnamefont {P.}~\bibnamefont
  {Kramer}}\ and\ \bibinfo {author} {\bibfnamefont {M.}~\bibnamefont
  {Saraceno}},\ }\href {\doibase 10.1007/3-540-10579-4} {\emph {\bibinfo
  {title} {{Geometry of the Time-Dependent Variational Principle in Quantum
  Mechanics}}}}\ (\bibinfo  {publisher} {Springer},\ \bibinfo {address}
  {Berlin, Germany},\ \bibinfo {year} {1981})\BibitemShut {NoStop}%
\bibitem [{\citenamefont {Hackl}\ \emph {et~al.}(2020)\citenamefont {Hackl},
  \citenamefont {Guaita}, \citenamefont {Shi}, \citenamefont {Haegeman},
  \citenamefont {Demler},\ and\ \citenamefont {Cirac}}]{Hackl:2020/sp/048}%
  \BibitemOpen
  \bibfield  {author} {\bibinfo {author} {\bibfnamefont {L.}~\bibnamefont
  {Hackl}}, \bibinfo {author} {\bibfnamefont {T.}~\bibnamefont {Guaita}},
  \bibinfo {author} {\bibfnamefont {T.}~\bibnamefont {Shi}}, \bibinfo {author}
  {\bibfnamefont {J.}~\bibnamefont {Haegeman}}, \bibinfo {author}
  {\bibfnamefont {E.}~\bibnamefont {Demler}}, \ and\ \bibinfo {author}
  {\bibfnamefont {I.}~\bibnamefont {Cirac}},\ }\href {\doibase
  10.21468/SciPostPhys.9.4.048} {\bibfield  {journal} {\bibinfo  {journal}
  {SciPost Phys.}\ }\textbf {\bibinfo {volume} {9}},\ \bibinfo {pages} {048}
  (\bibinfo {year} {2020})}\BibitemShut {NoStop}%
\bibitem [{\citenamefont {Caillat}\ \emph {et~al.}(2005)\citenamefont
  {Caillat}, \citenamefont {Zanghellini}, \citenamefont {Kitzler},
  \citenamefont {Koch}, \citenamefont {Kreuzer},\ and\ \citenamefont
  {Scrinzi}}]{Caillat:2005/pra/012712}%
  \BibitemOpen
  \bibfield  {author} {\bibinfo {author} {\bibfnamefont {J.}~\bibnamefont
  {Caillat}}, \bibinfo {author} {\bibfnamefont {J.}~\bibnamefont
  {Zanghellini}}, \bibinfo {author} {\bibfnamefont {M.}~\bibnamefont
  {Kitzler}}, \bibinfo {author} {\bibfnamefont {O.}~\bibnamefont {Koch}},
  \bibinfo {author} {\bibfnamefont {W.}~\bibnamefont {Kreuzer}}, \ and\
  \bibinfo {author} {\bibfnamefont {A.}~\bibnamefont {Scrinzi}},\ }\href
  {\doibase 10.1103/PhysRevA.71.012712} {\bibfield  {journal} {\bibinfo
  {journal} {Phys. Rev. A}\ }\textbf {\bibinfo {volume} {71}},\ \bibinfo
  {pages} {012712} (\bibinfo {year} {2005})}\BibitemShut {NoStop}%
\bibitem [{\citenamefont {Sasmal}\ and\ \citenamefont
  {Vendrell}(2020)}]{Sasmal:2020/jcp/154110}%
  \BibitemOpen
  \bibfield  {author} {\bibinfo {author} {\bibfnamefont {S.}~\bibnamefont
  {Sasmal}}\ and\ \bibinfo {author} {\bibfnamefont {O.}~\bibnamefont
  {Vendrell}},\ }\href {\doibase 10.1063/5.0028116} {\bibfield  {journal}
  {\bibinfo  {journal} {J. Chem. Phys.}\ }\textbf {\bibinfo {volume} {153}},\
  \bibinfo {pages} {154110} (\bibinfo {year} {2020})}\BibitemShut {NoStop}%
\bibitem [{\citenamefont {Beck}\ \emph {et~al.}(2000)\citenamefont {Beck},
  \citenamefont {J\"{a}ckle}, \citenamefont {Worth},\ and\ \citenamefont
  {Meyer}}]{Beck:2000/pr/1}%
  \BibitemOpen
  \bibfield  {author} {\bibinfo {author} {\bibfnamefont {M.}~\bibnamefont
  {Beck}}, \bibinfo {author} {\bibfnamefont {A.}~\bibnamefont {J\"{a}ckle}},
  \bibinfo {author} {\bibfnamefont {G.}~\bibnamefont {Worth}}, \ and\ \bibinfo
  {author} {\bibfnamefont {H.-D.}\ \bibnamefont {Meyer}},\ }\href {\doibase
  10.1016/S0370-1573(99)00047-2} {\bibfield  {journal} {\bibinfo  {journal}
  {Phys. Rep.}\ }\textbf {\bibinfo {volume} {324}},\ \bibinfo {pages} {1}
  (\bibinfo {year} {2000})}\BibitemShut {NoStop}%
\bibitem [{\citenamefont {Meyer}, \citenamefont {Gatti},\ and\ \citenamefont
  {Worth}(2009)}]{Meyer:2009/book}%
  \BibitemOpen
  \bibinfo {editor} {\bibfnamefont {H.-D.}\ \bibnamefont {Meyer}}, \bibinfo
  {editor} {\bibfnamefont {F.}~\bibnamefont {Gatti}}, \ and\ \bibinfo {editor}
  {\bibfnamefont {G.~A.}\ \bibnamefont {Worth}},\ eds.,\ \href@noop {} {\emph
  {\bibinfo {title} {{Multidimensional Quantum Dynamics: MCTDH Theory and
  Applications}}}}\ (\bibinfo  {publisher} {Wiley-VCH},\ \bibinfo {address}
  {Weinheim},\ \bibinfo {year} {2009})\BibitemShut {NoStop}%
\bibitem [{\citenamefont {Burghardt}, \citenamefont {Meyer},\ and\
  \citenamefont {Cederbaum}(1999)}]{Burghardt:1999/jcp/2927}%
  \BibitemOpen
  \bibfield  {author} {\bibinfo {author} {\bibfnamefont {I.}~\bibnamefont
  {Burghardt}}, \bibinfo {author} {\bibfnamefont {H.-D.}\ \bibnamefont
  {Meyer}}, \ and\ \bibinfo {author} {\bibfnamefont {L.~S.}\ \bibnamefont
  {Cederbaum}},\ }\href {\doibase 10.1063/1.479574} {\bibfield  {journal}
  {\bibinfo  {journal} {J. Chem. Phys.}\ }\textbf {\bibinfo {volume} {111}},\
  \bibinfo {pages} {2927} (\bibinfo {year} {1999})}\BibitemShut {NoStop}%
\bibitem [{\citenamefont {Worth}, \citenamefont {Robb},\ and\ \citenamefont
  {Burghardt}(2004)}]{Worth:2004/fd/307}%
  \BibitemOpen
  \bibfield  {author} {\bibinfo {author} {\bibfnamefont {G.~A.}\ \bibnamefont
  {Worth}}, \bibinfo {author} {\bibfnamefont {M.~A.}\ \bibnamefont {Robb}}, \
  and\ \bibinfo {author} {\bibfnamefont {I.}~\bibnamefont {Burghardt}},\ }\href
  {\doibase 10.1039/B314253A} {\bibfield  {journal} {\bibinfo  {journal}
  {Farad. Discuss.}\ }\textbf {\bibinfo {volume} {127}},\ \bibinfo {pages}
  {307} (\bibinfo {year} {2004})}\BibitemShut {NoStop}%
\bibitem [{\citenamefont {Richings}\ \emph {et~al.}(2015)\citenamefont
  {Richings}, \citenamefont {Polyak}, \citenamefont {Spinlove}, \citenamefont
  {Worth}, \citenamefont {Burghardt},\ and\ \citenamefont
  {Lasorne}}]{Richings:2015/irpc/269}%
  \BibitemOpen
  \bibfield  {author} {\bibinfo {author} {\bibfnamefont {G.~W.}\ \bibnamefont
  {Richings}}, \bibinfo {author} {\bibfnamefont {I.}~\bibnamefont {Polyak}},
  \bibinfo {author} {\bibfnamefont {K.~E.}\ \bibnamefont {Spinlove}}, \bibinfo
  {author} {\bibfnamefont {G.~A.}\ \bibnamefont {Worth}}, \bibinfo {author}
  {\bibfnamefont {I.}~\bibnamefont {Burghardt}}, \ and\ \bibinfo {author}
  {\bibfnamefont {B.}~\bibnamefont {Lasorne}},\ }\href {\doibase
  10.1080/0144235X.2015.1051354} {\bibfield  {journal} {\bibinfo  {journal}
  {Int. Rev. Phys. Chem.}\ }\textbf {\bibinfo {volume} {34}},\ \bibinfo {pages}
  {269} (\bibinfo {year} {2015})}\BibitemShut {NoStop}%
\bibitem [{\citenamefont {Joubert-Doriol}\ and\ \citenamefont
  {Izmaylov}(2018)}]{Joubert:2018/jpca/6031}%
  \BibitemOpen
  \bibfield  {author} {\bibinfo {author} {\bibfnamefont {L.}~\bibnamefont
  {Joubert-Doriol}}\ and\ \bibinfo {author} {\bibfnamefont {A.~F.}\
  \bibnamefont {Izmaylov}},\ }\href {\doibase 10.1021/acs.jpca.8b03404}
  {\bibfield  {journal} {\bibinfo  {journal} {J. Phys. Chem. A}\ }\textbf
  {\bibinfo {volume} {122}},\ \bibinfo {pages} {6031} (\bibinfo {year}
  {2018})}\BibitemShut {NoStop}%
\bibitem [{\citenamefont {Worth}(2020)}]{Worth:2020/cpc/107040}%
  \BibitemOpen
  \bibfield  {author} {\bibinfo {author} {\bibfnamefont {G.~A.}\ \bibnamefont
  {Worth}},\ }\href {\doibase 10.1016/j.cpc.2019.107040} {\bibfield  {journal}
  {\bibinfo  {journal} {Comput. Phys. Commun.}\ }\textbf {\bibinfo {volume}
  {248}},\ \bibinfo {pages} {107040} (\bibinfo {year} {2020})}\BibitemShut
  {NoStop}%
\bibitem [{\citenamefont {Ben-Nun}, \citenamefont {Quenneville},\ and\
  \citenamefont {Martinez}(2000)}]{Bennun:2000/jpca/5161}%
  \BibitemOpen
  \bibfield  {author} {\bibinfo {author} {\bibfnamefont {M.}~\bibnamefont
  {Ben-Nun}}, \bibinfo {author} {\bibfnamefont {J.}~\bibnamefont
  {Quenneville}}, \ and\ \bibinfo {author} {\bibfnamefont {T.~J.}\ \bibnamefont
  {Martinez}},\ }\href {\doibase 10.1021/jp994174i} {\bibfield  {journal}
  {\bibinfo  {journal} {J. Phys. Chem. A}\ }\textbf {\bibinfo {volume} {104}},\
  \bibinfo {pages} {5161} (\bibinfo {year} {2000})}\BibitemShut {NoStop}%
\bibitem [{\citenamefont {Ben-Nun}\ and\ \citenamefont
  {Mart{\ifmmode\acute{\imath}\else\'{\i}\fi}nez}(2002)}]{Ben-Nun:2002/Book}%
  \BibitemOpen
  \bibfield  {author} {\bibinfo {author} {\bibfnamefont {M.}~\bibnamefont
  {Ben-Nun}}\ and\ \bibinfo {author} {\bibfnamefont {{\relax Todd}.~J.}\
  \bibnamefont {Mart{\ifmmode\acute{\imath}\else\'{\i}\fi}nez}},\ }in\ \href
  {\doibase 10.1002/0471264318.ch7} {\emph {\bibinfo {booktitle} {{Advances in
  Chemical Physics}}}}\ (\bibinfo  {publisher} {John Wiley {\&} Sons, Inc.},\
  \bibinfo {address} {Hoboken, NJ, USA},\ \bibinfo {year} {2002})\ pp.\
  \bibinfo {pages} {439--512}\BibitemShut {NoStop}%
\bibitem [{\citenamefont {Makhov}\ \emph {et~al.}(2017)\citenamefont {Makhov},
  \citenamefont {Symonds}, \citenamefont {Fernandez-Alberti},\ and\
  \citenamefont {Shalashilin}}]{Makhov:2017/cp/200}%
  \BibitemOpen
  \bibfield  {author} {\bibinfo {author} {\bibfnamefont {D.~V.}\ \bibnamefont
  {Makhov}}, \bibinfo {author} {\bibfnamefont {C.}~\bibnamefont {Symonds}},
  \bibinfo {author} {\bibfnamefont {S.}~\bibnamefont {Fernandez-Alberti}}, \
  and\ \bibinfo {author} {\bibfnamefont {D.~V.}\ \bibnamefont {Shalashilin}},\
  }\href {\doibase 10.1016/j.chemphys.2017.04.003} {\bibfield  {journal}
  {\bibinfo  {journal} {Chem. Phys.}\ }\textbf {\bibinfo {volume} {493}},\
  \bibinfo {pages} {200} (\bibinfo {year} {2017})}\BibitemShut {NoStop}%
\bibitem [{\citenamefont {Curchod}\ and\ \citenamefont
  {Mart{\ifmmode\acute{\imath}\else\'{\i}\fi}nez}(2018)}]{Curchod:2018/cr/3305}%
  \BibitemOpen
  \bibfield  {author} {\bibinfo {author} {\bibfnamefont {B.}~\bibnamefont
  {Curchod}}\ and\ \bibinfo {author} {\bibfnamefont {T.~J.}\ \bibnamefont
  {Mart{\ifmmode\acute{\imath}\else\'{\i}\fi}nez}},\ }\href {\doibase
  10.1021/acs.chemrev.7b00423} {\bibfield  {journal} {\bibinfo  {journal}
  {Chem. Rev.}\ }\textbf {\bibinfo {volume} {118}},\ \bibinfo {pages} {3305}
  (\bibinfo {year} {2018})}\BibitemShut {NoStop}%
\bibitem [{\citenamefont {Lassmann}\ and\ \citenamefont
  {Curchod}(2021)}]{Lassmann:2021/jcp/211106}%
  \BibitemOpen
  \bibfield  {author} {\bibinfo {author} {\bibfnamefont {Y.}~\bibnamefont
  {Lassmann}}\ and\ \bibinfo {author} {\bibfnamefont {B.~F.~E.}\ \bibnamefont
  {Curchod}},\ }\href {\doibase 10.1063/5.0052118} {\bibfield  {journal}
  {\bibinfo  {journal} {J. Chem. Phys.}\ }\textbf {\bibinfo {volume} {154}},\
  \bibinfo {pages} {211106} (\bibinfo {year} {2021})}\BibitemShut {NoStop}%
\bibitem [{\citenamefont {Heller}(1975)}]{Heller:1975/jcp/1544}%
  \BibitemOpen
  \bibfield  {author} {\bibinfo {author} {\bibfnamefont {E.~J.}\ \bibnamefont
  {Heller}},\ }\href {\doibase 10.1063/1.430620} {\bibfield  {journal}
  {\bibinfo  {journal} {J. Chem. Phys.}\ }\textbf {\bibinfo {volume} {62}},\
  \bibinfo {pages} {1544} (\bibinfo {year} {1975})}\BibitemShut {NoStop}%
\bibitem [{\citenamefont {Heller}(1976)}]{Heller:1976/jcp/63}%
  \BibitemOpen
  \bibfield  {author} {\bibinfo {author} {\bibfnamefont {E.~J.}\ \bibnamefont
  {Heller}},\ }\href {\doibase 10.1063/1.431911} {\bibfield  {journal}
  {\bibinfo  {journal} {J. Chem. Phys.}\ }\textbf {\bibinfo {volume} {64}},\
  \bibinfo {pages} {63} (\bibinfo {year} {1976})}\BibitemShut {NoStop}%
\bibitem [{\citenamefont {Lasser}\ and\ \citenamefont
  {Lubich}(2020)}]{Lasser:2020/an/229}%
  \BibitemOpen
  \bibfield  {author} {\bibinfo {author} {\bibfnamefont {C.}~\bibnamefont
  {Lasser}}\ and\ \bibinfo {author} {\bibfnamefont {C.}~\bibnamefont
  {Lubich}},\ }\href {\doibase 10.1017/S0962492920000033} {\bibfield  {journal}
  {\bibinfo  {journal} {Acta Numer.}\ }\textbf {\bibinfo {volume} {29}},\
  \bibinfo {pages} {229} (\bibinfo {year} {2020})}\BibitemShut {NoStop}%
\bibitem [{\citenamefont {Garashchuk}(2021)}]{Garashchuk:2021/Book}%
  \BibitemOpen
  \bibfield  {author} {\bibinfo {author} {\bibfnamefont {S.}~\bibnamefont
  {Garashchuk}},\ }in\ \href {\doibase 10.1007/978-3-030-67262-1_8} {\emph
  {\bibinfo {booktitle} {{Basis Sets in Computational Chemistry}}}}\ (\bibinfo
  {publisher} {Springer},\ \bibinfo {address} {Cham, Switzerland},\ \bibinfo
  {year} {2021})\ pp.\ \bibinfo {pages} {215--252}\BibitemShut {NoStop}%
\bibitem [{\citenamefont {Bargmann}\ \emph {et~al.}(1971)\citenamefont
  {Bargmann}, \citenamefont {Butera}, \citenamefont {Girardello},\ and\
  \citenamefont {Klauder}}]{Bargmann:1971/rmp/221}%
  \BibitemOpen
  \bibfield  {author} {\bibinfo {author} {\bibfnamefont {V.}~\bibnamefont
  {Bargmann}}, \bibinfo {author} {\bibfnamefont {P.}~\bibnamefont {Butera}},
  \bibinfo {author} {\bibfnamefont {L.}~\bibnamefont {Girardello}}, \ and\
  \bibinfo {author} {\bibfnamefont {J.~R.}\ \bibnamefont {Klauder}},\ }\href
  {\doibase 10.1016/0034-4877(71)90006-1} {\bibfield  {journal} {\bibinfo
  {journal} {Rep. Math. Phys.}\ }\textbf {\bibinfo {volume} {2}},\ \bibinfo
  {pages} {221} (\bibinfo {year} {1971})}\BibitemShut {NoStop}%
\bibitem [{\citenamefont {Shalashilin}\ and\ \citenamefont
  {Child}(2001)}]{Shalashilin:2001/jcp/5367}%
  \BibitemOpen
  \bibfield  {author} {\bibinfo {author} {\bibfnamefont {D.~V.}\ \bibnamefont
  {Shalashilin}}\ and\ \bibinfo {author} {\bibfnamefont {M.~S.}\ \bibnamefont
  {Child}},\ }\href {\doibase 10.1063/1.1394939} {\bibfield  {journal}
  {\bibinfo  {journal} {J. Chem. Phys.}\ }\textbf {\bibinfo {volume} {115}},\
  \bibinfo {pages} {5367} (\bibinfo {year} {2001})}\BibitemShut {NoStop}%
\bibitem [{\citenamefont {Werther}, \citenamefont {Choudhury},\ and\
  \citenamefont {Gro{\ss}mann}(2021)}]{Werther:2021/irpc/81}%
  \BibitemOpen
  \bibfield  {author} {\bibinfo {author} {\bibfnamefont {M.}~\bibnamefont
  {Werther}}, \bibinfo {author} {\bibfnamefont {S.~L.}\ \bibnamefont
  {Choudhury}}, \ and\ \bibinfo {author} {\bibfnamefont {F.}~\bibnamefont
  {Gro{\ss}mann}},\ }\href {\doibase 10.1080/0144235X.2020.1823168} {\bibfield
  {journal} {\bibinfo  {journal} {Int. Rev. Phys. Chem.}\ }\textbf {\bibinfo
  {volume} {40}},\ \bibinfo {pages} {81} (\bibinfo {year} {2021})}\BibitemShut
  {NoStop}%
\bibitem [{\citenamefont {Shalashilin}(2009)}]{Shalashilin:2009/jcp/244101}%
  \BibitemOpen
  \bibfield  {author} {\bibinfo {author} {\bibfnamefont {D.~V.}\ \bibnamefont
  {Shalashilin}},\ }\href {\doibase 10.1063/1.3153302} {\bibfield  {journal}
  {\bibinfo  {journal} {J. Chem. Phys.}\ }\textbf {\bibinfo {volume} {130}},\
  \bibinfo {pages} {244101} (\bibinfo {year} {2009})}\BibitemShut {NoStop}%
\bibitem [{\citenamefont {Makhov}\ \emph {et~al.}(2022)\citenamefont {Makhov},
  \citenamefont {Adeyemi}, \citenamefont {Cowperthwaite},\ and\ \citenamefont
  {Shalashilin}}]{Makhov:2022/jpc/025001}%
  \BibitemOpen
  \bibfield  {author} {\bibinfo {author} {\bibfnamefont {D.~V.}\ \bibnamefont
  {Makhov}}, \bibinfo {author} {\bibfnamefont {S.}~\bibnamefont {Adeyemi}},
  \bibinfo {author} {\bibfnamefont {M.}~\bibnamefont {Cowperthwaite}}, \ and\
  \bibinfo {author} {\bibfnamefont {D.~V.}\ \bibnamefont {Shalashilin}},\
  }\href {\doibase 10.1088/2399-6528/ac4d39} {\bibfield  {journal} {\bibinfo
  {journal} {J. Phys. Commun.}\ }\textbf {\bibinfo {volume} {6}},\ \bibinfo
  {pages} {025001} (\bibinfo {year} {2022})}\BibitemShut {NoStop}%
\bibitem [{\citenamefont {Yu}\ \emph {et~al.}(2020)\citenamefont {Yu},
  \citenamefont {Bannwarth}, \citenamefont {Liang}, \citenamefont
  {Hohenstein},\ and\ \citenamefont
  {Mart{\ifmmode\acute{\imath}\else\'{\i}\fi}nez}}]{Yu:2020/jacs/20680}%
  \BibitemOpen
  \bibfield  {author} {\bibinfo {author} {\bibfnamefont {J.~K.}\ \bibnamefont
  {Yu}}, \bibinfo {author} {\bibfnamefont {C.}~\bibnamefont {Bannwarth}},
  \bibinfo {author} {\bibfnamefont {R.}~\bibnamefont {Liang}}, \bibinfo
  {author} {\bibfnamefont {E.~G.}\ \bibnamefont {Hohenstein}}, \ and\ \bibinfo
  {author} {\bibfnamefont {T.~J.}\ \bibnamefont
  {Mart{\ifmmode\acute{\imath}\else\'{\i}\fi}nez}},\ }\href {\doibase
  10.1021/jacs.0c09056} {\bibfield  {journal} {\bibinfo  {journal} {J. Am.
  Chem. Soc.}\ }\textbf {\bibinfo {volume} {142}},\ \bibinfo {pages} {20680}
  (\bibinfo {year} {2020})}\BibitemShut {NoStop}%
\bibitem [{\citenamefont {Ibele}\ and\ \citenamefont
  {Curchod}(2021)}]{Ibele:2021/jcp/174119}%
  \BibitemOpen
  \bibfield  {author} {\bibinfo {author} {\bibfnamefont {L.~M.}\ \bibnamefont
  {Ibele}}\ and\ \bibinfo {author} {\bibfnamefont {B.~F.~E.}\ \bibnamefont
  {Curchod}},\ }\href {\doibase 10.1063/5.0071376} {\bibfield  {journal}
  {\bibinfo  {journal} {J. Chem. Phys.}\ }\textbf {\bibinfo {volume} {155}},\
  \bibinfo {pages} {174119} (\bibinfo {year} {2021})}\BibitemShut {NoStop}%
\bibitem [{\citenamefont {Sawada}\ \emph {et~al.}(1985)\citenamefont {Sawada},
  \citenamefont {Heather}, \citenamefont {Jackson},\ and\ \citenamefont
  {Metiu}}]{Sawada:1985/jcp/3009}%
  \BibitemOpen
  \bibfield  {author} {\bibinfo {author} {\bibfnamefont {S.-I.}\ \bibnamefont
  {Sawada}}, \bibinfo {author} {\bibfnamefont {R.}~\bibnamefont {Heather}},
  \bibinfo {author} {\bibfnamefont {B.}~\bibnamefont {Jackson}}, \ and\
  \bibinfo {author} {\bibfnamefont {H.}~\bibnamefont {Metiu}},\ }\href
  {\doibase 10.1063/1.449204} {\bibfield  {journal} {\bibinfo  {journal} {J.
  Chem. Phys.}\ }\textbf {\bibinfo {volume} {83}},\ \bibinfo {pages} {3009}
  (\bibinfo {year} {1985})}\BibitemShut {NoStop}%
\bibitem [{\citenamefont {Kay}(1989)}]{Kay:1989/cp/165}%
  \BibitemOpen
  \bibfield  {author} {\bibinfo {author} {\bibfnamefont {K.~G.}\ \bibnamefont
  {Kay}},\ }\href {\doibase 10.1016/0301-0104(89)87102-2} {\bibfield  {journal}
  {\bibinfo  {journal} {Chem. Phys.}\ }\textbf {\bibinfo {volume} {137}},\
  \bibinfo {pages} {165} (\bibinfo {year} {1989})}\BibitemShut {NoStop}%
\bibitem [{\citenamefont {Habershon}(2012)}]{Habershon:2012/jcp/014109}%
  \BibitemOpen
  \bibfield  {author} {\bibinfo {author} {\bibfnamefont {S.}~\bibnamefont
  {Habershon}},\ }\href {\doibase 10.1063/1.3671978} {\bibfield  {journal}
  {\bibinfo  {journal} {J. Chem. Phys.}\ }\textbf {\bibinfo {volume} {136}},\
  \bibinfo {pages} {014109} (\bibinfo {year} {2012})}\BibitemShut {NoStop}%
\bibitem [{\citenamefont {Manthe}(2015)}]{Manthe:2015/jcp/244109}%
  \BibitemOpen
  \bibfield  {author} {\bibinfo {author} {\bibfnamefont {U.}~\bibnamefont
  {Manthe}},\ }\href {\doibase 10.1063/1.4922889} {\bibfield  {journal}
  {\bibinfo  {journal} {J. Chem. Phys.}\ }\textbf {\bibinfo {volume} {142}},\
  \bibinfo {pages} {244109} (\bibinfo {year} {2015})}\BibitemShut {NoStop}%
\bibitem [{\citenamefont {Lubich}(2015)}]{Lubich:2015/amre/311}%
  \BibitemOpen
  \bibfield  {author} {\bibinfo {author} {\bibfnamefont {C.}~\bibnamefont
  {Lubich}},\ }\href {\doibase 10.1093/amrx/abv006} {\bibfield  {journal}
  {\bibinfo  {journal} {Appl. Math. Res. EXpress}\ }\textbf {\bibinfo {volume}
  {2015}},\ \bibinfo {pages} {311} (\bibinfo {year} {2015})}\BibitemShut
  {NoStop}%
\bibitem [{\citenamefont {Meyer}\ and\ \citenamefont
  {Wang}(2018)}]{Meyer:2018/jcp/124105}%
  \BibitemOpen
  \bibfield  {author} {\bibinfo {author} {\bibfnamefont {H.-D.}\ \bibnamefont
  {Meyer}}\ and\ \bibinfo {author} {\bibfnamefont {H.}~\bibnamefont {Wang}},\
  }\href {\doibase 10.1063/1.5024859} {\bibfield  {journal} {\bibinfo
  {journal} {J. Chem. Phys.}\ }\textbf {\bibinfo {volume} {148}},\ \bibinfo
  {pages} {124105} (\bibinfo {year} {2018})}\BibitemShut {NoStop}%
\bibitem [{\citenamefont {Polyak}, \citenamefont {Allan},\ and\ \citenamefont
  {Worth}(2015)}]{Polyak:2015/jcp/084121}%
  \BibitemOpen
  \bibfield  {author} {\bibinfo {author} {\bibfnamefont {I.}~\bibnamefont
  {Polyak}}, \bibinfo {author} {\bibfnamefont {C.~S.~M.}\ \bibnamefont
  {Allan}}, \ and\ \bibinfo {author} {\bibfnamefont {G.~A.}\ \bibnamefont
  {Worth}},\ }\href {\doibase 10.1063/1.4929478} {\bibfield  {journal}
  {\bibinfo  {journal} {J. Chem. Phys.}\ }\textbf {\bibinfo {volume} {143}},\
  \bibinfo {pages} {084121} (\bibinfo {year} {2015})}\BibitemShut {NoStop}%
\bibitem [{\citenamefont {Wu}\ and\ \citenamefont
  {Batista}(2003)}]{Wu:2003/jcp/6720}%
  \BibitemOpen
  \bibfield  {author} {\bibinfo {author} {\bibfnamefont {Y.}~\bibnamefont
  {Wu}}\ and\ \bibinfo {author} {\bibfnamefont {V.~S.}\ \bibnamefont
  {Batista}},\ }\href {\doibase 10.1063/1.1560636} {\bibfield  {journal}
  {\bibinfo  {journal} {J. Chem. Phys.}\ }\textbf {\bibinfo {volume} {118}},\
  \bibinfo {pages} {6720} (\bibinfo {year} {2003})}\BibitemShut {NoStop}%
\bibitem [{\citenamefont {Koch}\ and\ \citenamefont
  {Frankcombe}(2013)}]{Koch:2013/prl/263202}%
  \BibitemOpen
  \bibfield  {author} {\bibinfo {author} {\bibfnamefont {W.}~\bibnamefont
  {Koch}}\ and\ \bibinfo {author} {\bibfnamefont {T.~J.}\ \bibnamefont
  {Frankcombe}},\ }\href {\doibase 10.1103/PhysRevLett.110.263202} {\bibfield
  {journal} {\bibinfo  {journal} {Phys. Rev. Lett.}\ }\textbf {\bibinfo
  {volume} {110}},\ \bibinfo {pages} {263202} (\bibinfo {year}
  {2013})}\BibitemShut {NoStop}%
\bibitem [{\citenamefont {Meyer}, \citenamefont {Manthe},\ and\ \citenamefont
  {Cederbaum}(1990)}]{Meyer:1990/pcl/73}%
  \BibitemOpen
  \bibfield  {author} {\bibinfo {author} {\bibfnamefont {H.-D.}\ \bibnamefont
  {Meyer}}, \bibinfo {author} {\bibfnamefont {U.}~\bibnamefont {Manthe}}, \
  and\ \bibinfo {author} {\bibfnamefont {L.~S.}\ \bibnamefont {Cederbaum}},\
  }\href {\doibase 10.1016/0009-2614(90)87014-I} {\bibfield  {journal}
  {\bibinfo  {journal} {Chem. Phys. Lett.}\ }\textbf {\bibinfo {volume}
  {165}},\ \bibinfo {pages} {73} (\bibinfo {year} {1990})}\BibitemShut
  {NoStop}%
\bibitem [{\citenamefont {Manthe}, \citenamefont {Meyer},\ and\ \citenamefont
  {Cederbaum}(1992)}]{Manthe:1992/jcp/3199}%
  \BibitemOpen
  \bibfield  {author} {\bibinfo {author} {\bibfnamefont {U.}~\bibnamefont
  {Manthe}}, \bibinfo {author} {\bibfnamefont {H.-D.}\ \bibnamefont {Meyer}}, \
  and\ \bibinfo {author} {\bibfnamefont {L.~S.}\ \bibnamefont {Cederbaum}},\
  }\href {\doibase 10.1063/1.463007} {\bibfield  {journal} {\bibinfo  {journal}
  {J. Chem. Phys.}\ }\textbf {\bibinfo {volume} {97}},\ \bibinfo {pages} {3199}
  (\bibinfo {year} {1992})}\BibitemShut {NoStop}%
\bibitem [{\citenamefont {Martinez}, \citenamefont {Ben-Nun},\ and\
  \citenamefont {Ashkenazi}(1996)}]{Martinez:1996/jcp/2847}%
  \BibitemOpen
  \bibfield  {author} {\bibinfo {author} {\bibfnamefont {T.~J.}\ \bibnamefont
  {Martinez}}, \bibinfo {author} {\bibfnamefont {M.}~\bibnamefont {Ben-Nun}}, \
  and\ \bibinfo {author} {\bibfnamefont {G.}~\bibnamefont {Ashkenazi}},\ }\href
  {\doibase 10.1063/1.471108} {\bibfield  {journal} {\bibinfo  {journal} {J.
  Chem. Phys.}\ }\textbf {\bibinfo {volume} {104}},\ \bibinfo {pages} {2847}
  (\bibinfo {year} {1996})}\BibitemShut {NoStop}%
\bibitem [{\citenamefont
  {Mauritz~Andersson}(2001)}]{MauritzAndersson:2001/jcp/1158}%
  \BibitemOpen
  \bibfield  {author} {\bibinfo {author} {\bibfnamefont {L.}~\bibnamefont
  {Mauritz~Andersson}},\ }\href {\doibase 10.1063/1.1380204} {\bibfield
  {journal} {\bibinfo  {journal} {J. Chem. Phys.}\ }\textbf {\bibinfo {volume}
  {115}},\ \bibinfo {pages} {1158} (\bibinfo {year} {2001})}\BibitemShut
  {NoStop}%
\bibitem [{\citenamefont {Conte}\ and\ \citenamefont
  {Lubich}(2010)}]{Conte:2010/m2an/759}%
  \BibitemOpen
  \bibfield  {author} {\bibinfo {author} {\bibfnamefont {D.}~\bibnamefont
  {Conte}}\ and\ \bibinfo {author} {\bibfnamefont {C.}~\bibnamefont {Lubich}},\
  }\href {\doibase 10.1051/m2an/2010018} {\bibfield  {journal} {\bibinfo
  {journal} {ESAIM: M2AN}\ }\textbf {\bibinfo {volume} {44}},\ \bibinfo {pages}
  {759} (\bibinfo {year} {2010})}\BibitemShut {NoStop}%
\bibitem [{\citenamefont {Werther}\ and\ \citenamefont
  {Gro{\ss}mann}(2020)}]{Werther:2020/prb/174315}%
  \BibitemOpen
  \bibfield  {author} {\bibinfo {author} {\bibfnamefont {M.}~\bibnamefont
  {Werther}}\ and\ \bibinfo {author} {\bibfnamefont {F.}~\bibnamefont
  {Gro{\ss}mann}},\ }\href {\doibase 10.1103/PhysRevB.101.174315} {\bibfield
  {journal} {\bibinfo  {journal} {Phys. Rev. B}\ }\textbf {\bibinfo {volume}
  {101}},\ \bibinfo {pages} {174315} (\bibinfo {year} {2020})}\BibitemShut
  {NoStop}%
\bibitem [{\citenamefont {Makhov}\ \emph {et~al.}(2014)\citenamefont {Makhov},
  \citenamefont {Glover}, \citenamefont {Martinez},\ and\ \citenamefont
  {Shalashilin}}]{Makhov:2014/jcp/054110}%
  \BibitemOpen
  \bibfield  {author} {\bibinfo {author} {\bibfnamefont {D.~V.}\ \bibnamefont
  {Makhov}}, \bibinfo {author} {\bibfnamefont {W.~J.}\ \bibnamefont {Glover}},
  \bibinfo {author} {\bibfnamefont {T.~J.}\ \bibnamefont {Martinez}}, \ and\
  \bibinfo {author} {\bibfnamefont {D.~V.}\ \bibnamefont {Shalashilin}},\
  }\href {\doibase 10.1063/1.4891530} {\bibfield  {journal} {\bibinfo
  {journal} {J. Chem. Phys.}\ }\textbf {\bibinfo {volume} {141}},\ \bibinfo
  {pages} {054110} (\bibinfo {year} {2014})}\BibitemShut {NoStop}%
\bibitem [{\citenamefont {Izmaylov}(2013)}]{Izmaylov:2013/jcp/104115}%
  \BibitemOpen
  \bibfield  {author} {\bibinfo {author} {\bibfnamefont {A.~F.}\ \bibnamefont
  {Izmaylov}},\ }\href {\doibase 10.1063/1.4794047} {\bibfield  {journal}
  {\bibinfo  {journal} {J. Chem. Phys.}\ }\textbf {\bibinfo {volume} {138}},\
  \bibinfo {pages} {104115} (\bibinfo {year} {2013})}\BibitemShut {NoStop}%
\bibitem [{\citenamefont {Izmaylov}\ and\ \citenamefont
  {Joubert-Doriol}(2017)}]{Izmaylov:2017/jpcl/1793}%
  \BibitemOpen
  \bibfield  {author} {\bibinfo {author} {\bibfnamefont {A.~F.}\ \bibnamefont
  {Izmaylov}}\ and\ \bibinfo {author} {\bibfnamefont {L.}~\bibnamefont
  {Joubert-Doriol}},\ }\href {\doibase 10.1021/acs.jpclett.7b00596} {\bibfield
  {journal} {\bibinfo  {journal} {J. Phys. Chem. Lett.}\ }\textbf {\bibinfo
  {volume} {8}},\ \bibinfo {pages} {1793} (\bibinfo {year} {2017})}\BibitemShut
  {NoStop}%
\bibitem [{\citenamefont {Martinazzo}\ and\ \citenamefont
  {Burghardt}(2020)}]{Martinazzo:2020/prl/150601}%
  \BibitemOpen
  \bibfield  {author} {\bibinfo {author} {\bibfnamefont {R.}~\bibnamefont
  {Martinazzo}}\ and\ \bibinfo {author} {\bibfnamefont {I.}~\bibnamefont
  {Burghardt}},\ }\href {\doibase 10.1103/PhysRevLett.124.150601} {\bibfield
  {journal} {\bibinfo  {journal} {Phys. Rev. Lett.}\ }\textbf {\bibinfo
  {volume} {124}},\ \bibinfo {pages} {150601} (\bibinfo {year}
  {2020})}\BibitemShut {NoStop}%
\bibitem [{\citenamefont {Mendive-Tapia}\ and\ \citenamefont
  {Meyer}(2020)}]{Mendive-Tapia:2020/jcp/234114}%
  \BibitemOpen
  \bibfield  {author} {\bibinfo {author} {\bibfnamefont {D.}~\bibnamefont
  {Mendive-Tapia}}\ and\ \bibinfo {author} {\bibfnamefont {H.-D.}\ \bibnamefont
  {Meyer}},\ }\href {\doibase 10.1063/5.0035581} {\bibfield  {journal}
  {\bibinfo  {journal} {J. Chem. Phys.}\ }\textbf {\bibinfo {volume} {153}},\
  \bibinfo {pages} {234114} (\bibinfo {year} {2020})}\BibitemShut {NoStop}%
\bibitem [{Note1()}]{Note1}%
  \BibitemOpen
  \bibinfo {note} {The complex Stiefel manifold is the space of all
  $k$-dimensional semi-unitary matrices in a $D$-dimensional space with
  $k<D$.}\BibitemShut {Stop}%
\bibitem [{\citenamefont {Hairer}, \citenamefont {Wanner},\ and\ \citenamefont
  {Lubich}(2006)}]{Hairer:2006/Book}%
  \BibitemOpen
  \bibfield  {author} {\bibinfo {author} {\bibfnamefont {E.}~\bibnamefont
  {Hairer}}, \bibinfo {author} {\bibfnamefont {G.}~\bibnamefont {Wanner}}, \
  and\ \bibinfo {author} {\bibfnamefont {C.}~\bibnamefont {Lubich}},\ }\href
  {\doibase 10.1007/3-540-30666-8} {\emph {\bibinfo {title} {{Geometric
  Numerical Integration}}}}\ (\bibinfo  {publisher} {Springer},\ \bibinfo
  {address} {Berlin, Germany},\ \bibinfo {year} {2006})\BibitemShut {NoStop}%
\bibitem [{\citenamefont {Gower}\ and\ \citenamefont
  {Dijksterhuis}(2004)}]{Gower:2004/Book}%
  \BibitemOpen
  \bibfield  {author} {\bibinfo {author} {\bibfnamefont {J.~C.}\ \bibnamefont
  {Gower}}\ and\ \bibinfo {author} {\bibfnamefont {G.~B.}\ \bibnamefont
  {Dijksterhuis}},\ }in\ \href {\doibase
  10.1093/acprof:oso/9780198510581.001.0001} {\emph {\bibinfo {booktitle}
  {{Procrustes Problems}}}}\ (\bibinfo  {publisher} {Oxford University Press},\
  \bibinfo {address} {Oxford, England, UK},\ \bibinfo {year}
  {2004})\BibitemShut {NoStop}%
\bibitem [{\citenamefont {Maskri}\ and\ \citenamefont
  {Joubert-Doriol}(2022)}]{Maskri:2022/ptrsa/20200379}%
  \BibitemOpen
  \bibfield  {author} {\bibinfo {author} {\bibfnamefont {R.}~\bibnamefont
  {Maskri}}\ and\ \bibinfo {author} {\bibfnamefont {L.}~\bibnamefont
  {Joubert-Doriol}},\ }\href {\doibase 10.1098/rsta.2020.0379} {\bibfield
  {journal} {\bibinfo  {journal} {Phil. Trans. R. Soc. A.}\ }\textbf {\bibinfo
  {volume} {380}},\ \bibinfo {pages} {20200379} (\bibinfo {year}
  {2022})}\BibitemShut {NoStop}%
\bibitem [{\citenamefont {Eaton}\ \emph {et~al.}(2017)\citenamefont {Eaton},
  \citenamefont {Bateman}, \citenamefont {Hauberg},\ and\ \citenamefont
  {Wehbring}}]{octave}%
  \BibitemOpen
  \bibfield  {author} {\bibinfo {author} {\bibfnamefont {J.~W.}\ \bibnamefont
  {Eaton}}, \bibinfo {author} {\bibfnamefont {D.}~\bibnamefont {Bateman}},
  \bibinfo {author} {\bibfnamefont {S.}~\bibnamefont {Hauberg}}, \ and\
  \bibinfo {author} {\bibfnamefont {R.}~\bibnamefont {Wehbring}},\ }\href
  {https://www.gnu.org/software/octave/doc/v4.2.1/} {\emph {\bibinfo {title}
  {{GNU Octave} version 4.2.1 manual: a high-level interactive language for
  numerical computations}}} (\bibinfo {year} {2017})\BibitemShut {NoStop}%
\bibitem [{\citenamefont {Joubert-Doriol}\ and\ \citenamefont
  {Izmaylov}(2015)}]{Joubert:2015/jcp/134107}%
  \BibitemOpen
  \bibfield  {author} {\bibinfo {author} {\bibfnamefont {L.}~\bibnamefont
  {Joubert-Doriol}}\ and\ \bibinfo {author} {\bibfnamefont {A.~F.}\
  \bibnamefont {Izmaylov}},\ }\href {\doibase 10.1063/1.4916384} {\bibfield
  {journal} {\bibinfo  {journal} {J. Chem. Phys.}\ }\textbf {\bibinfo {volume}
  {142}},\ \bibinfo {pages} {134107} (\bibinfo {year} {2015})}\BibitemShut
  {NoStop}%
\bibitem [{\citenamefont {R{\ifmmode\ddot{o}\else\"{o}\fi}mer}, \citenamefont
  {Ruckenbauer},\ and\ \citenamefont
  {Burghardt}(2013)}]{Romer:2013/jcp/064106}%
  \BibitemOpen
  \bibfield  {author} {\bibinfo {author} {\bibfnamefont {S.}~\bibnamefont
  {R{\ifmmode\ddot{o}\else\"{o}\fi}mer}}, \bibinfo {author} {\bibfnamefont
  {M.}~\bibnamefont {Ruckenbauer}}, \ and\ \bibinfo {author} {\bibfnamefont
  {I.}~\bibnamefont {Burghardt}},\ }\href {\doibase 10.1063/1.4788830}
  {\bibfield  {journal} {\bibinfo  {journal} {J. Chem. Phys.}\ }\textbf
  {\bibinfo {volume} {138}},\ \bibinfo {pages} {064106} (\bibinfo {year}
  {2013})}\BibitemShut {NoStop}%
\bibitem [{\citenamefont {Eisenbrandt}\ \emph {et~al.}(2018)\citenamefont
  {Eisenbrandt}, \citenamefont {Ruckenbauer}, \citenamefont
  {R{\ifmmode\ddot{o}\else\"{o}\fi}mer},\ and\ \citenamefont
  {Burghardt}}]{Eisenbrandt:2018/jcp/174101}%
  \BibitemOpen
  \bibfield  {author} {\bibinfo {author} {\bibfnamefont {P.}~\bibnamefont
  {Eisenbrandt}}, \bibinfo {author} {\bibfnamefont {M.}~\bibnamefont
  {Ruckenbauer}}, \bibinfo {author} {\bibfnamefont {S.}~\bibnamefont
  {R{\ifmmode\ddot{o}\else\"{o}\fi}mer}}, \ and\ \bibinfo {author}
  {\bibfnamefont {I.}~\bibnamefont {Burghardt}},\ }\href {\doibase
  10.1063/1.5053414} {\bibfield  {journal} {\bibinfo  {journal} {J. Chem.
  Phys.}\ }\textbf {\bibinfo {volume} {149}},\ \bibinfo {pages} {174101}
  (\bibinfo {year} {2018})}\BibitemShut {NoStop}%
\end{thebibliography}%
\end{document}